\documentclass[12pt]{article}
\usepackage{amsmath}
\usepackage{latexsym}
\usepackage{float}
\usepackage{amssymb}
\usepackage{graphicx}
\begin{document}

\title{Structure of Polymer Brushes in Cylindrical Tubes: A
Molecular Dynamics Simulation}

\author{Dimitar I. Dimitrov$^1$, Andrey Milchev$^{2,3}$, Kurt Binder$^3$, and 
Dieter W. Heermann$^4$\\
$^{1}$ {Inorganic Chemistry and Physical Chemistry
Department, University}\\ 
{of Food Technology, 26 Maritza Blvd., 4002 Plovdiv, Bulgaria} \\
$^{2}$ {Institute of Physical Chemistry, Bulgarian Academy of
Sciences,}\\ 
{1113 Sofia, Bulgaria} \\
$^{3}$ {Institut f\"ur Physik, Johannes Gutenberg Universit\"at
Mainz,}\\ 
{Staudinger Weg 7, 55099 Mainz, Germany} \\
$^{4}$ {Institut f\"ur Theoretische Physik, Universit\"at Heidelberg,}\\
{Philosophenweg 19, 69120 Heidelberg, Germany}
}

\maketitle


\begin{abstract}
Molecular Dynamics simulations of a coarse-grained bead-spring
model of flexible macromolecules tethered with one end to the
surface of a cylindrical pore are presented. Chain length $N$ and
grafting density $\sigma$ are varied over a wide range and the
crossover from ``mushroom'' to ``brush'' behavior is studied for
three pore diameters. The monomer density profile and the
distribution of the free chain ends are computed and compared to
the corresponding model of polymer brushes at flat substrates. It
is found that there exists a regime of $N$ and $\sigma$ for large
enough pore diameter where the brush height in the pore exceeds
the brush height on the flat substrate, while for large enough $N$
and $\sigma$ (and small enough pore diameters) the opposite
behavior occurs, i.e. the brush is compressed by confinement.
These findings are used to discuss the corresponding theories on
polymer brushes at concave substrates.
\end{abstract}

\section{Introduction}
By special endgroups flexible macromolecules can be tethered to a
substrate via its chain end. In a dilute solution under good
solvent conditions using a non-adsorbing substrate surface that
repels the monomers one thus obtains a ``mushroom'' configuration
of the polymer \cite{1,2}. If one grafts many such polymer chains
on the substrate, such that the polymer coils in their mushroom
configuration would strongly overlap each other, excluded volume
interactions reduce this overlap by leading to strongly stretched
configurations of the coils, the so-called ``polymer brushes''
\cite{2,3}. In this way, the substrate is coated with a polymeric
layer of height $h$ proportional to the chain length $N$ of these
endgrafted polymers \cite{2,3,4,5}. Of course, the distinction between 
"mushrooms" and "brushes" is not a completely sharp one, but rather these states
 of a surface coated by endgrafted polymers are separated by broad crossover regime
 where the coils overlap weakly and get more and more stretched as with increasing 
grafting density the overlap increases.

Since such polymer brushes find important applications such as
colloid stabilization, lubrication, creation of adhesive layers,
or control of wetting properties, etc. \cite{6}, there has been an
enormous interest in these systems, ranging from various methods
of their synthesis \cite{6} to their theoretical understanding by
analytic theory \cite{2} - \cite{4}, \cite{7} - \cite{25} and computer simulation
 \cite{26} - \cite{39}. Most of this work
concerns polymers tethered to flat planar substrates, however, and
brushes on curved substrates were considered only occasionally
\cite{10,17,21,22,23,24,25,30,35,36,39} although this case clearly is
relevant in practice, too. Since it was stated that brushes
adsorbed on the outside of spherical or cylindrical surfaces differ in
many interesting aspects from flat brushes \cite{17}, while
brushes at the inside of spheres or cylinders were claimed to differ relatively
little from the planar substrate case \cite{17}, not much
attention was paid to brushes grafted to the surface of spherical
\cite{22,25,35} or cylindrical \cite{10,22,25,39} pores. While it
is clear that such a brush gets compressed when the pore diameter
$D$ becomes comparable to the height $h$, the brush would adopt
under otherwise identical conditions, conflicting predictions
appear in the literature concerning the behavior for large $D$:
while the Sevick theory \cite{22} suggests that the brush height
is maximal for $D\rightarrow \infty$ and decreases monotonously
with $1/D$, self-consistent field arguments \cite{10,25} imply
that for small $1/D$ the brush expands slightly. No simulations
have so far been carried out for brushes in cylindrical pores yet
at all, apart form a study of electroosmotic flow in capillaries
and its control by endgrafted chains \cite{39} (this work was
restricted to a single chain length $N=10$ and the mushroom
regime, however).

In the present paper, we wish to fill this gap by presenting
Molecular Dynamics simulations of polymer brushes in cylindrical
pores, and hence present numerical evidence to settle this issue
whether brushes on weakly concave substrates shrink or expand, in
comparison with the flat case.

\section{Model and Simulation Method}
The present study does not concern a chemically realistic
description of a specific polymer, but rather we are interested in
the generic features of polymer brushes grafted inside of
cylindrical tubes only. Therefore we disregard any details of
chemical structure, and work with a coarse-grained bead-spring
model only. The beads hence represent effective monomeric units,
as usual, like Kuhn segments, so the nearest-neighbor distance
along a bead-spring chain does not correspond to a chemical bond
between carbon atoms ($\approx 0.15 nm$) but rather is of the
order of $1 nm$. On this scale, the effects of the torsional and
bond-angle potentials can be considered to be averaged out, and
all what is left is an effective interaction between monomeric
units and an effective spring potential between neighboring
monomeric units (beads) along the chain \cite{40,41,42,43,44}.
There is ample evidence that such models are useful to describe
many properties of macromolecular systems \cite{40,41,42,43,44},
and in fact previous work on simulation of polymer brushes on flat
substrates \cite{26,27,31,34,36,38,45} and on the outside of
nanocylinders \cite{30} have used such models successfully.

The interactions between the beads are then modeled by a
Lennard-Jones (LJ) potential $U_{LJ}(r)$ that is truncated at
$r_c$ and shifted to zero there to avoid a singularity in the
force,
\begin{equation}\label{eq1}
U(\vec{r}_i-\vec{r}_j)=U_{LJ}(r)-U_{LJ}(r_c),\; r =
|\vec{r}_i-\vec{r}_j|,
\end{equation}
\begin{equation}\label{eq2}
U_{LJ}(r)=4 \epsilon[(\sigma_{mm}/r)^{12}-(\sigma_{mm}/r)^6],
\end{equation}
where $\vec{r}_i$ is the position of the i'th bead, the parameters
$\epsilon$ and $\sigma_{mm}$ set the scales for the strength and
range of the $LJ$ potential, and the cutoff distance $r_c$ is
chosen as
\begin{equation}\label{eq3}
r_c=2^{7/6}\sigma_{mm}\;.
\end{equation}
Note that most previous work on polymer brushes (apart from
\cite{31,38,45}) have truncated and shifted the potential right in
its minimum, i.e. choosing $r_c=2^{1/6} \sigma _{mm}$ rather than
Eq.~(\ref{eq3}), and then the potential is purely repulsive. Such
an almost athermal model can describe a polymer brush under
conditions of a very good solvent only, but it would not be
possible to consider brushes under conditions of a bad solvent or
under Theta-conditions \cite{46}. For the model defined in
Eqs.~(\ref{eq1}), (\ref{eq2}), (\ref{eq3}) the Theta temperature
$\Theta$ has been estimated as \cite{47} $k_B\Theta/\epsilon \approx
3.3$. In the present work, we study polymer brushes under
conditions of a (moderately) good solvent only, choosing the
temperature of the simulation throughout as $k_BT/\epsilon = 4.0$,
following \cite{38}. The spring potential is created by adding a
finitely extensible nonlinear elastic (FENE) potential \cite{48}
\begin{equation}\label{eq4}
U_{FENE}(r)=-15 \epsilon (R_0/\sigma_{mm})^2 \ln
(1-r^2/R_0^2)\;,\quad R_0 = 1.5 \sigma _{mm},
\end{equation}
to Eq.~(\ref{eq1}) if monomers $i,j$ are nearest neighbors along a
chain. The choice of the additional parameters that appear in
Eq.~(\ref{eq4}) ensure that there is no tendency for the beads to
form a simple crystal structure (face-centered cubic or
body-centered cubic, for instance) at high density, because the
minimum of the potential for two bonded monomers (neighbors along
a chain) occurs at $r_{min}\approx 0.96 \sigma_{mm}$, rather
different from (and incommensurate with) the minimum distance of
$U_{LJ}(r)$, $r_{min} \approx 1.12 \sigma_{mm}$. Therefore the
present model at low temperatures, $T$ far below $\Theta$, freezes
into a glass-like structure, both for ordinary melts \cite{49}
(i.e., chains that are not grafted to a substrate) and for polymer
brushes \cite{50}. However, studies of the present model of a
polymer brush in a cylinder at and below the Theta temperature
will be left to future work.

The grafting wall is represented by particles forming a triangular
lattice, wrapped on a torus by a periodic boundary condition, and
choosing the triangular lattice spacing and the number of wall
particles such that a cylinder of diameter $D$ results. The
particles forming this cylindrical wall are bound together by the
same potential as the spring potential between the beads of the
chain, defined by Eqs.~(\ref{eq1})-(\ref{eq4}). Thus it is ensured
(through the repulsive part of the potential, Eq.~(\ref{eq2}))
that no bead can cross the cylinder surface, at the temperatures
of interest, to get out of the cylinder. In the direction along
the straight cylinder axis, the standard periodic boundary condition is
taken, as usual. The grafted chain ends of the polymers are put on
regular positions of a lattice on the surface of a cylinder whose
radius is one length unit smaller than the cylinder forming the
confining tube.

This choice of cylindrical confinement is computationally
convenient and in fact was motivated by the research on carbon
nanotubes \cite{51}. However, it should be noted that by
techniques such as combinations of electron beam lithography and
nanoimprint lithography artificial nanochannels can be fabricated
with a width between $35 nm$ and $150 nm$ \cite{52}, and such
nanochannels were recently used to study statics and dynamics of
confined DNA macromolecules \cite{53}. Thus a different modeling
of the walls of the confining tube may be appropriate for such
artificial nanochannels, but this is left to later work. Still a
different modeling may be required to describe confinement of
grafted biopolymers in cylindrical pores in biological membranes,
however.

Particular care is needed to create an initial configuration of
the grafted polymers and equilibrate it. Periodically arranged
grafting sites are chosen simply to be the wall particles,
according to a chosen grafting density $\sigma$. Choosing our
units of length and temperature such that $\sigma _{mm}=1,
\epsilon=1,k_B \equiv 1$, we work with cylindrical diameters
$D=30,42$, and 62, the corresponding lengths of the cylinders in
the direction of the cylinder axis being $L=81.61,76.18$, and
67.47, respectively. For comparison, also brushes on flat substrates were studied
(formally this corresponds to $D = \infty$), keeping otherwise the
model parameters identical. Due to the difficulty of
equilibration, for $D=30$ the maximum chain length that could be
studied was $N=64$, for $D=42$ we could go up to $N=80$, and for
$D=60$ up to $N=112$, at the smallest grafting density studied,
and $N=24,32$ and 64, at the largest grafting densities used for
$D=30$, 42 and 62, respectively. At the highest grafting densities
studied, the cylindrical tubes were already filled uniformly with
monomeric units at melt density. If the grafting density is lower
than its maximal value, or the chain length shorter, an empty
region free of monomeric units remains present near the axis of
the cylindrical channel (Fig.~\ref{fig1}), and this free volume
facilitates equilibration.

Next we describe the construction of the initial configuration of
the grafted polymers. This construction begins for a given $(N,
\sigma,R$) with the growth of chains in a tube with radius
$R_{init}=2R$ and $N=N_{max}$, where $N_{max}$ is the number of
monomers in the longest chain investigated for this choice of
$\sigma$ and $R$. After the first monomer has been placed at the
grafting site, each subsequent monomer is put at a point at
distance $0.075 R_0$ from the previous one, along a direction
$\vec{r}=2.0* \vec{e}_{0z}+ \vec{e}_{rand}$, $\vec{e}_{0z}$
being a unit vector pointing radially from the grafting site to
the tube axis (i.e., along the z-axis), while $\vec{e}_{rand}$ is
another unit vector oriented at random. After an initial
equilibration of this system for about 10$^5$ MD steps, the tube
radius is rescaled to the desired final value $R$, and again the
system is equilibrated for 10$^5$ MD steps. From the final
configuration of this system, several systems with different
values of N are created, by cutting of the monomers near the free
chain end consecutively and eliminating them. These systems then
are further equilibrated for about 3.10$^5$ MD steps. These then
are the starting configurations for the averaging process
(typically statistics is collected during runs lasting from
2.10$^5$ to 4.10$^5$ MD steps). Equilibration was done by
Molecular Dynamics (MD), applying the standard leapfrog algorithm
\cite{42}, keeping the temperature constant via the Nose-Hoover
thermostat \cite{42}. Choosing the mass of an effective monomer as
$m=1$, the MD time unit is $t_0=(\sigma^2_{mm}m/48
\epsilon)^{1/2}=1/\sqrt{48}$, and the integration time step then
is chosen as $\delta t=0.0005 \; t_0$. Always 10$^6$ MD steps were
discarded for equilibration before averages were taken. In this
paper, we focus on our results on the density profiles of the
effective monomers and of the free chain ends, while corresponding
data on the chain extensions parallel and perpendicular to the
tube axis will be analyzed elsewhere \cite{54}.

\section{Density Profiles and Chain End Distributions in Confined
Cylindrical Brushes}

Figs.~\ref{fig2} - \ref{fig5} show a representative selection of
our simulation results, displaying in the upper part the brush
density profile and in the lower part the corresponding radial
distribution of free chain ends, both for a number of chain
lengths $N$ for the chosen combination of parameters $D$ and
$\sigma$, respectively. This is
only a small subset of all the data that have been generated. 
Here $\phi (r)$ is the number density of monomers in a radial shell between $r$ and $r + dr$,
 normalized such that $2 \pi L \int \limits _0 ^\frac{D}{2} r dr \phi (r)$ yields the total number of
 monomers in the system, and the end monomer distribution $\rho (r)$ is defined accordingly.

It is seen that for small $N$, $(N \leq 16$), in all cases the
behavior of these distribution functions is typical of polymer
mushroom behavior, as it is recalled in part (a) of Fig.\ref{fig5}
for a polymer brush at a planar grafting surface, but choosing
$\sigma$ so small that the chains are essentially noninteracting.
Then one still finds a small-scale structure in $\phi(r)$, due to
the nearest and next nearest neighbors of the grafting site
(counted along the chain), which is absent in the end monomer
distribution, of course. But both $\phi (r)$ and $\rho(r)$ have a
broad peak of comparable width and a similar decay to zero at
large distances from the grafting site.

 When the chain length  $N$ (and | or the grafting density $\sigma$) gradually increases one observes a smooth
 crossover of the distributions $\phi (r)$ and $\rho (r)$ to a behavior that is typical for
 polymer brushes on flat substrates. 
The behavior is completely different for large chains and/or high
grafting densities: then the monomer density becomes uniform
throughout the cylinder, apart from the immediate neighborhood of
the grafting surface, and the end monomer distribution gets peaked
in the cylinder axis region. There is, however, a smooth - almost
linear - decay of this end monomer distribution towards the
grafting surface: thus the approximation of this distribution by a
delta function obviously is very inaccurate. When one compares
these density profiles in cylindrical tubes (Figs.~\ref{fig2} -
\ref{fig4}) to the corresponding density profiles of brushes at
flat substrates, one notes great similarity between both cases
near the grafting surfaces. In the outer region of the brushes,
however, the behavior is markedly different: for a brush on a flat
surface extending into the half space, near the outer boundary of
the brush (i.e., near $z=h$ where $h$ is the ``brush height''
\cite{3,4}) there is a slow and gradual decay of all profiles
towards zero. For the brush in the cylinder, however, this is not
the case: at $r=0$ both $\phi(r)$ and $\rho(r)$ reach nonzero
limits, for long enough chains and/or high enough grafting
densities. When we continue to call such a state of endgrafted polymers in
 a cylindrical tube a "brush" we do not imply anything about the stretching of the chains, of course.
In addition one must consider the possibility that
monomers of a chain are anywhere inside of the cylinder, and hence
they are not confined to the semi-cylinder where the grafting
point is located, but also in the opposite semi-cylinder. Thus
when we introduce an orthogonal coordinate system such that the
origin coincides with the grafting site of a chain, the $y$-axis
running within the cylinder surface parallel to the cylinder axis,
and the $z$-axis runs perpendicular towards the cylinder axis,
monomers of long enough chains (in particular the chain ends), may
have $z$-coordinates in the full range $0<z<D$ and not only in the
range corresponding to the hemicylinder containing the grafting
point at its bottom, $0<z<D/2$. In the cylindrical coordinate
system one chooses the center of the coordinate system in the
cylinder axis, however, and $r$ is the radial distance from the
cylinder axis. We have two inequivalent entries for the same $r$,
namely
\begin{equation}\label{eq5}
r'=D/2-r, \; {\textrm{for}}\; z<D/2,\; {\textrm{and}}\; r'=r+D/2,\;
{\textrm{for}} \; z>D/2
\end{equation}
The information which monomers are in the upper semicylinder and which 
are in the lower semicylinder is lost when one uses the simple cylindrical
coordinates,
as done in Figs.~\ref{fig2} - \ref{fig4}. However, we have
nevertheless given this information, since it is a standard
assumption of the theoretical treatments that chains are confined
to the semicylinder $z<D/2$, or even the stronger assumption is
made that chains are confined to a cone, the cone axes running
from the grafting site along the $z$-axis with the cone top being in
the cylinder axis \cite{22,25}, the opening angle of the cone
being controlled by the grafting density via an adaptation of the
Alexander - de Gennes blob picture \cite{1,2,3} to the cylindrical
geometry. Our results imply that such a picture is by far too
restrictive, however.

Figs.~\ref{fig6} - \ref{fig9} now give the corresponding
distributions of the monomer density $\phi (r')$ and the chain ends
$\rho (r')$ for a number of typical cases. It is clearly recognized
that the monomers are not restricted to the region $r' <D/2$, and
actually the distributions at the distance $r'=D/2$ corresponding
to the cylinder axis are perfectly smooth. In this coordinate
system where the full range $0 < r' < D$ is considered, the
distributions are again rather similar to the corresponding
profiles of brushes grafted to flat substrates $(D \rightarrow
\infty$).

The profiles $\phi (r')$ have now been used to compute the first
moment $\langle r' \rangle $ and from $\langle r' \rangle $ one can
estimate the brush height $h$ as $h= 8 \langle r' \rangle /3$ (Note
that the (somewhat arbitrary) factor 8/3 is motivated by using the parabolic density
profile predicted by the strong stretching limit of the
self-consistent field theory for brushes at flat substrates,
although the actual profiles are rather different, as
Figs.~\ref{fig6} - \ref{fig9} show. But we use this factor $8/3$ so that our results
 for $h$ in the case of flat brushes conform to the results that one can find in the literature.): 

\begin{equation}\label{eq6}
h=\frac{8}{3} \langle r' \rangle, \; \langle r' \rangle = \int \limits _0 ^D r'
\phi (r') dr'/\int \limits
_0 ^D \phi (r') dr'
\end{equation}
For brushes on flat substrates one has the well known property $h
\propto N$ for large enough $N$. For brushes on cylindrical
substrates with small enough $D$ the brushes are compressed and hence a weaker
increase is expected. However, one expects from the self-consistent field theory
that for not too large $N$ (or large enough $D$, respectively, such that
the center of the cylinder is still more or less free of monomers, cf.
Fig~\ref{fig1}) that the density profile of a brush decreases somewhat 
slower in a cylindrical brush than for the corresponding blobs in a brush 
on the flat substrate, under identical conditions.
For the "blobs" close to the cylinder axis, less volume is available than
for corresponding blobs in a brush on a flat substrate. As a consequence, a
correspondingly stronger stretching tendency for chains in a blob grafted to the
inner surface of the cylinder would be predicted. Our numerical results
for brushes in cylinders with large enough $D$  (Fig.~\ref{fig10}) are
comparable with this prediction and with numerical results obtained by the SCMF
 (self consistent mean field) theory of Szleifer and Carignano \cite{35}. Hence, the theoretical result of Sevick
\cite{22} that the brush height is longest for $D \rightarrow \infty$ and decreases
monotonously
while decreasing $D$ is not confirmed by our simulations. Of course, this trend
seen on Fig. \ref{fig10}
cannot hold indefinitely with increasing $N$. Of course, Sevick theory \cite{22} is based on
 the Alexander-de Gennes ansatz that  the end monomers are
 localized at $r = h$  (for a flat brush this implies a delta function density profile $\phi (r)$ ), and thus it is not
 surprising that this theory is inaccurate.

When all the free volume in the
center of the brush (Fig. \ref{fig1})
is filled up with monomers, compression of the brush sets in, and consequently
the monomer density gets enhanced
also in the region of the brush close to the cylinder walls. In this regime, we
expect that the height
of a brush on a flat substrate would be larger than the height of the brush,
confined into the cylinder, and this is nicely demonstrated for a narrow enough
tube in Fig.~\ref{fig10} too. Of course, when there is no longer any region free of monomers
 left near the center of the tube, the notions of "polymer brush" and "brush height" lose their meaning.
 In fact, rather than comparing polymer brushes one may encounter states where the chain confirmations
 rather resemble "cigars", i.e. they are elongated along the axis of the pore.

\section{Conclusions}
In this paper, a first exploratory investigation of the structure of a brush
confined in cylinders of various diameters by means of Molecular Dynamics
computer simulations has been presented, using a (coarse-grained) bead-spring 
model under conditions of moderately good solvent (but without taking the
solvent 
particles into account explicitely - the effect of the solvent thus is described
only with respect to
static thermodynamic and structural properties, while the effect on dynamic
properties, e.g. via
mediating the hydrodynamic interactions, is missing). The walls of the cylinder
are described
in terms of a perfectly rigid lattice of particles that prevent the escape of
any monomer towards the outside.
The parameters of the problem (chain length $N$, grafting density $\sigma$,
cylinder diameter $D$) are varied and a
comparison with the properties of the corresponding brush model on a flat
substrate is also made.

In analysing the monomer and chain end density, it is suggested that one needs
to consider normally the standard
radial density profiles $\phi (r), \rho(r)$, but one may consider a
complementary coordinate $r'$,
which uses the information where the anchor point of the chain is: We define
$r'=D/2-r$ when the
considered monomer is in the same "hemicylinder" as the anchor point, while
$r'=D/2+r$ when the monomer 
is in the other "hemicylinder" (cf. the analogous profile of chains grafted
inside a sphere:
taking the grafting site of a chain as the south pole, it makes a difference
whether the monomer is
in the southern or northern hemisphere). Then the profiles are smooth also at
$r'=D/2$, corresponding to $r=0$,
see Figs.~\ref{fig6} -~\ref{fig8}, and resemble that of a brush on a flat
substrate (Fig.~\ref{fig5}).

On the other hand, the standard radial density profile (Figs.~\ref{fig2}
-~\ref{fig4}) is rather different as soon as
the central part of the cylinder is no longer empty (Fig.~\ref{fig1}), since
then at $r=0$ the densities happen to be nonzero and the similarity with the case
of brushes on flat surfaces is lost. That difference is reminiscent to the
case of density profiles when two brushes on planar surfaces are brought together
to such a small distance $D$ that
they start to interpenetrate. One still can ask a question what it the density
profile of the monomers (or chain ends)
of the left brush also at distance $z>D/2$ and likewise for the monomers (or chain
ends) of the right brush (grafting sites near $z=D$ at distances $z<D/2$). This
information on the interdigitation of the chains from the two brushes is of
critical importance when one considers dynamical behavior (e.g. when shear is
applied). Likewise, we feel that Figs.~\ref{fig6} -~\ref{fig8} contain an important
structural information that will play a role when one discusses the flow of
particles through the cylinder along its axis ( we shall investigate this
problem in future work).

One aspect that has been discussed contradictorily in the literature is the
effect of cylindrical confinement on the height of the brush - while the theory
of Sevick predicted that the maximal brush height occurs for the flat substrate,
and with increasing inverse radius of cylinder $1/R$ the height of the brush
decreases monotonously other arguments suggest that for small $1/R$ the brush
height increases first and decreases later, when a regime of  compressed brush
is reached, where no longer any free volume near the center of the cylinder is
left. Our results are comparable with
the latter picture, altough the change of the brush profile (and the brush
height) is rather small for small $1/R$.

Our results for the profiles of the monomer density and the chain ends
(Figs.~\ref{fig6} -~\ref{fig8}) also imply that the monomers of a chain are
{\em not} confined to a conical sector of the cylinder with the cone top in
the cylinder axis, as suggested by simple generalizations of the Alexander-De
Gennes blob picture \cite{2,3} for flat brushes to the cylindrical geometry
\cite{16,25} (Fig.~\ref{fig11}). Presumably for entropic reasons,
 it is much more favorable if a
much larger part of the cylinder volume is shared by many grafted chains, rather
than dividing up the cylinder into separate conical "compartments"
each containing a single chain. This finding is not really a surprize since
the Alexander-De Gennes blob picture is known to be rather unreliable also for
brushes on flat substrates, see e.g. \cite{32,33}.

There are many directions into which our study could be generalized: Obvious
generalizations include the study of dynamic corelations and response to
external perturbations. Of particular interest, both in nanotechnology and for
biological application would be to consider the transport of various molecules
or nanoparticles. It is conceivable that such a transport is either easily
possible or very difficult or even blocked. We plan to consider such extensions
of our study in future works. Given the fact that arrays of cylindrical nanotubes with
various diameters can be fabricated and used for various experiments, it is also
hoped that the present study will stimulate continuing experimental work on
brushes in such cylindrical geometries.

\section{Acknowledgments}
Support from the Deutsche Forshungsgemeinschaft (DFG) under project No 436 BUL 113/130 is gratefully
acknowledged. On of us (D.I.D.) thanks the Max-Planck society for support via a Max-Planck Fellowship
during the time this paper was completed.

\clearpage
\centerline{\includegraphics[width=3.7in, angle=270]{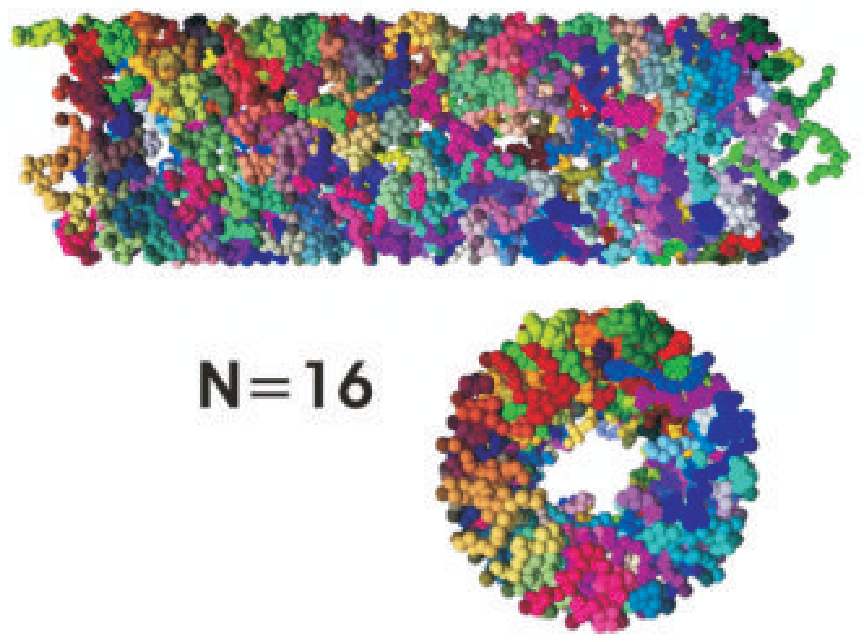}}
\vspace*{8pt}
\begin{figure}\caption{Snapshot picture of a brush grafted inside
of a cylinder, for $N=16$, $D=30, \sigma =0.08$, displaying different
chains in distinct colors in order to be able to distinguish them.
Top shows a side view of the cylinder, lower part a view of the
cross section. Note that the particles forming the cylindrical
wall are not displayed.\label{fig1}}
\end{figure}

\clearpage
\centerline{\includegraphics[width=2.8in, angle=270]{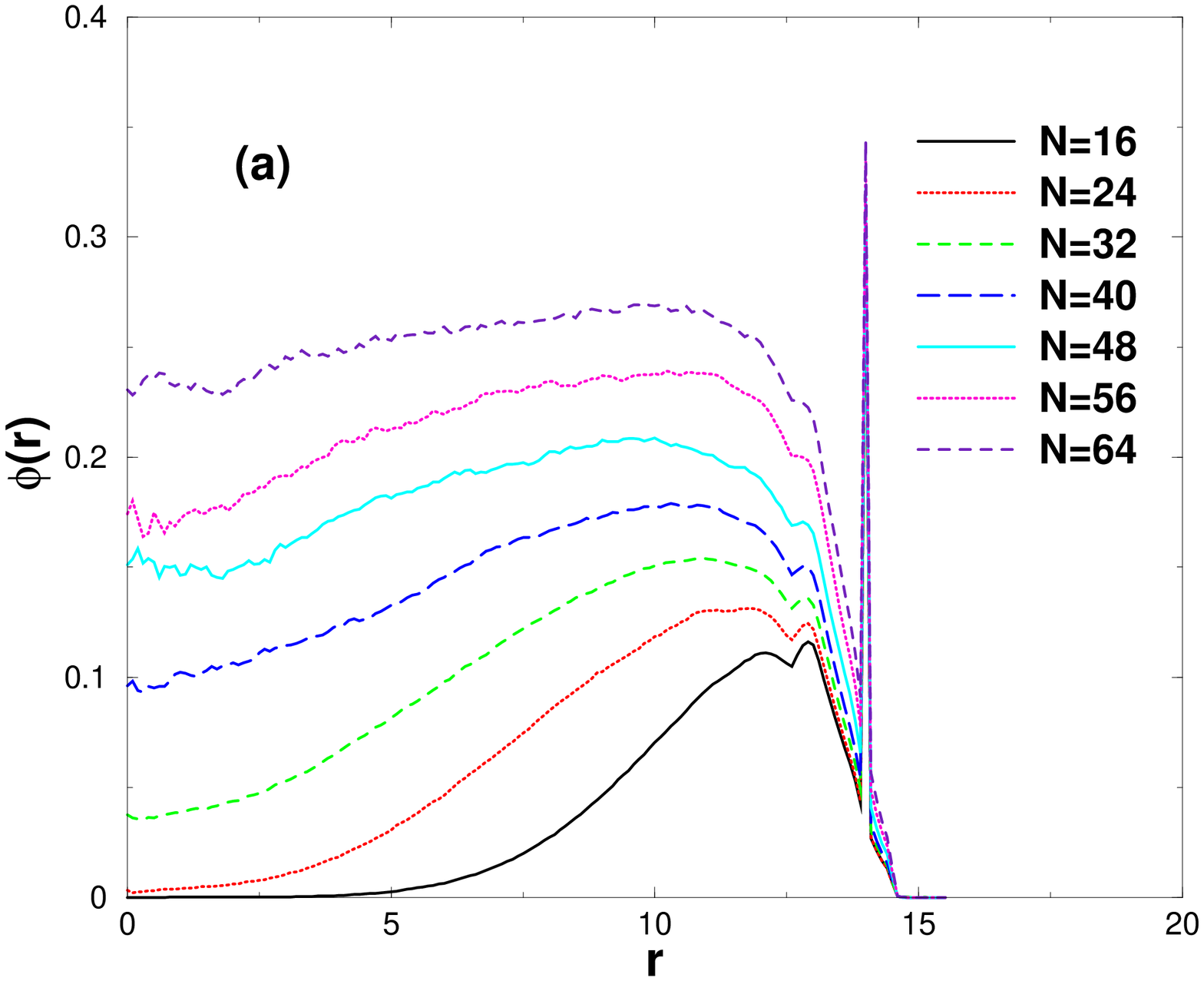}
\hspace*{50pt} \includegraphics[width=2.8in, angle=270]{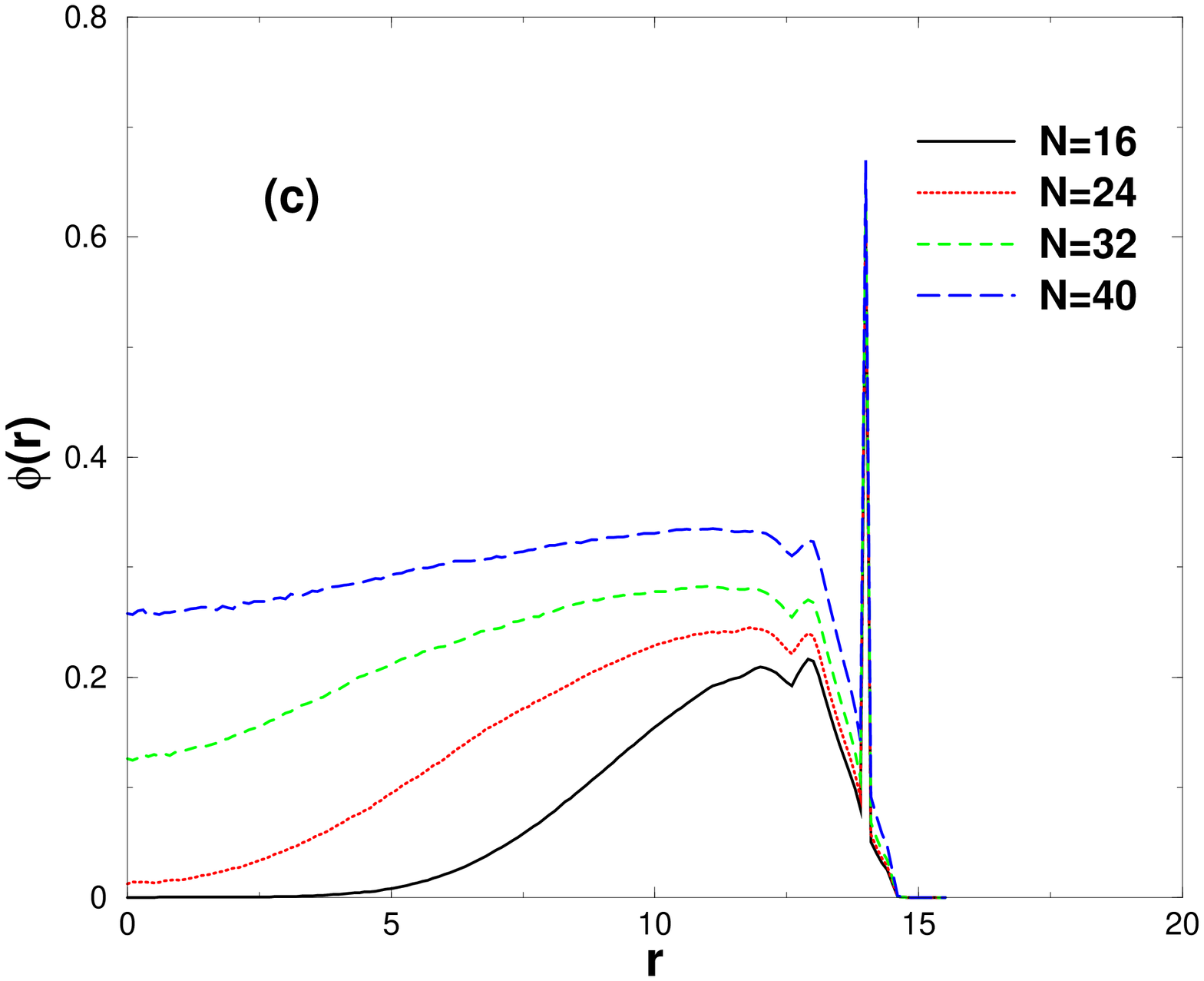}
}
\centerline{\includegraphics[width=2.8in, angle=270]{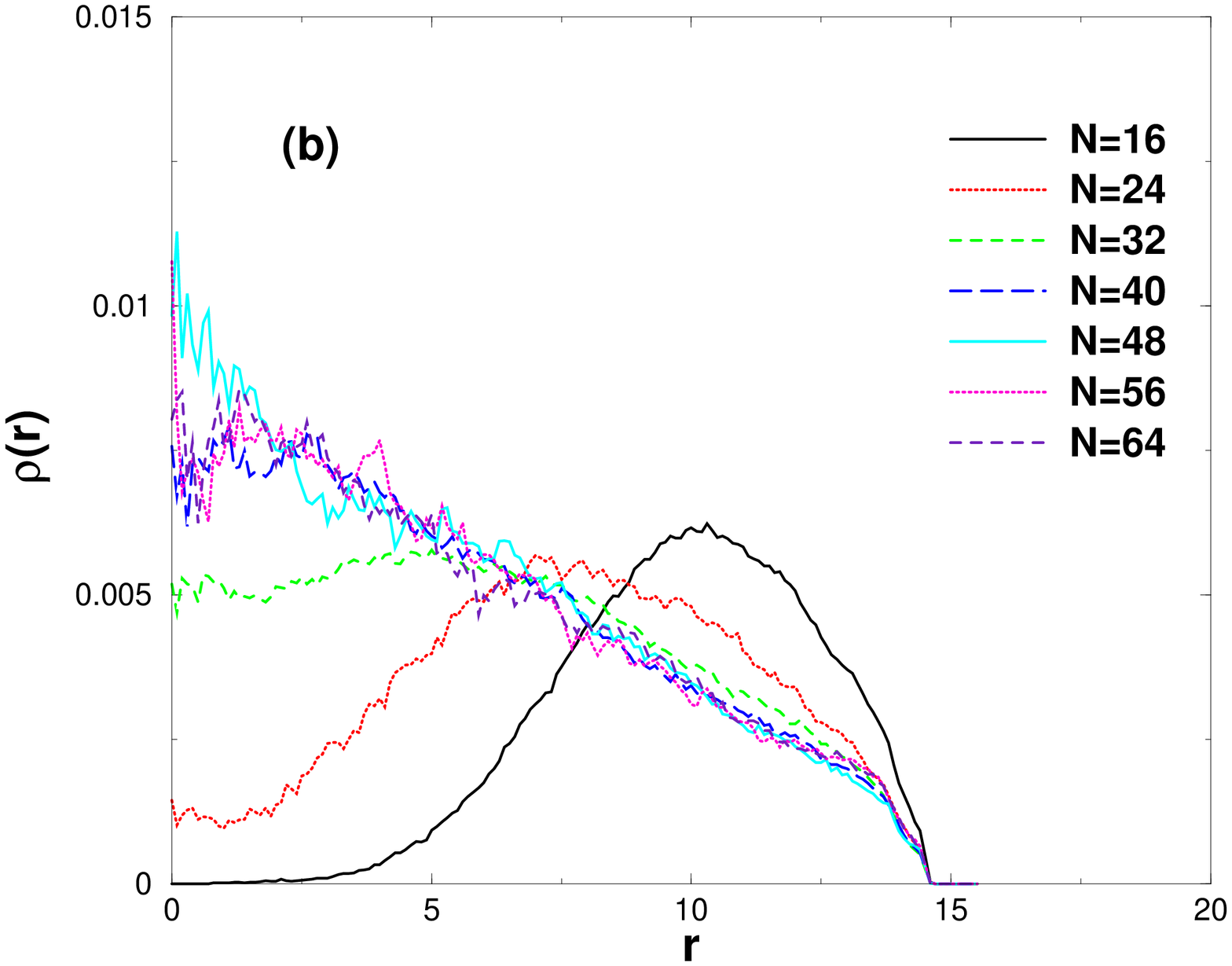}
\hspace*{50pt} \includegraphics[width=2.8in, angle=270]{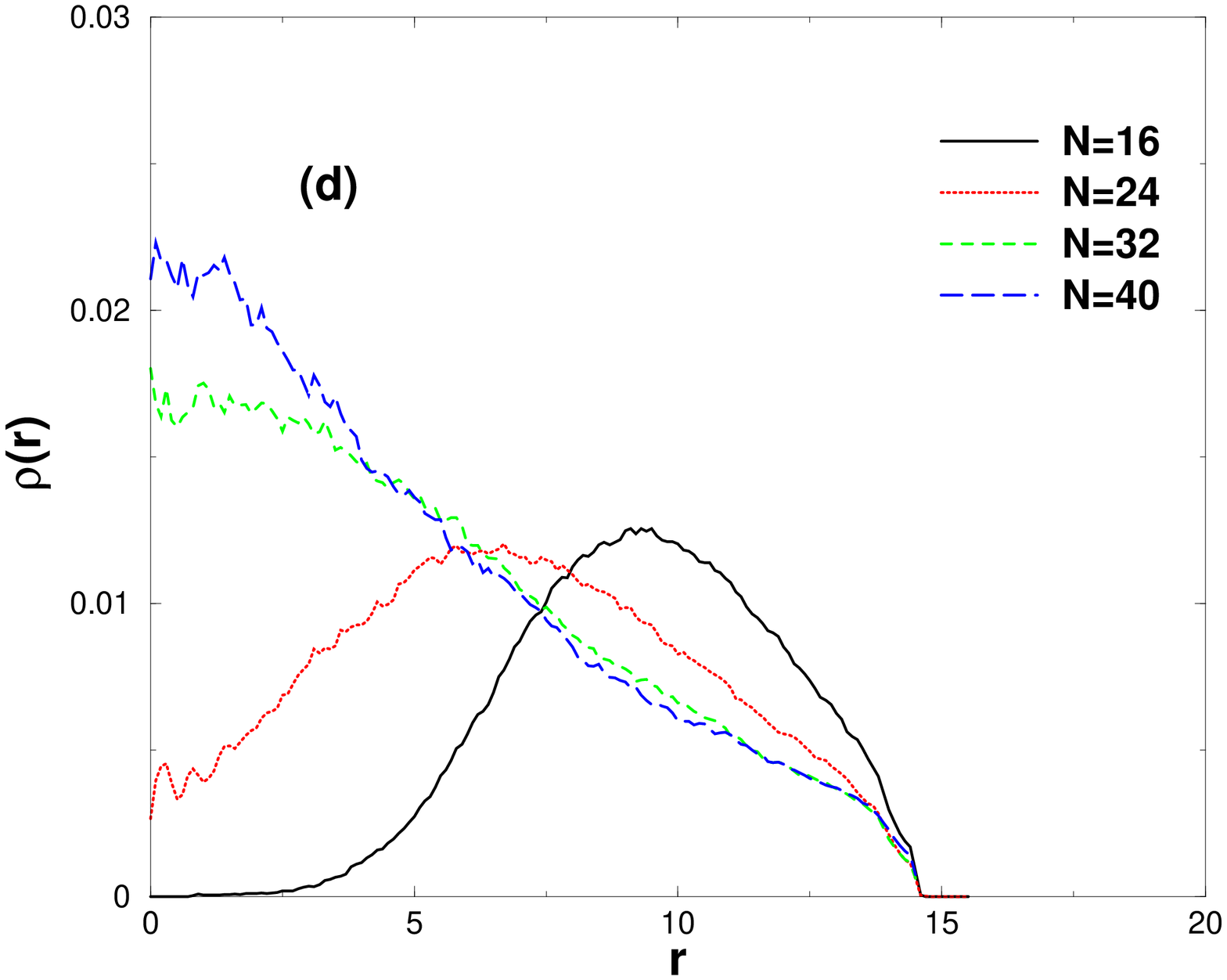}
}
\begin{figure}\caption{Radial distribution of the density $\phi
(r)$ of monomeric units (upper part) and of the free chain ends
$\rho (r)$ (lower part) plotted vs the radial distance $r$ from he
cylinder axis, for a cylindrical tube of diameter $D=30$. Note that
the sharp peak in $\phi (r)$ represents the first monomer of the
grafted chains at the fixed distance $\Delta r =1$ from the atoms
forming the cylindrical wall. Part (a,b) refers to
$N_{ch}=196\;(\sigma = 0.02730)$ and part (c,d) refers to
$N_{ch}=400 \; (\sigma = 0.05572)$.\label{fig2}}
\end{figure}

\clearpage
\begin{figure}
\centerline{\includegraphics[width=2.8in, angle=270]{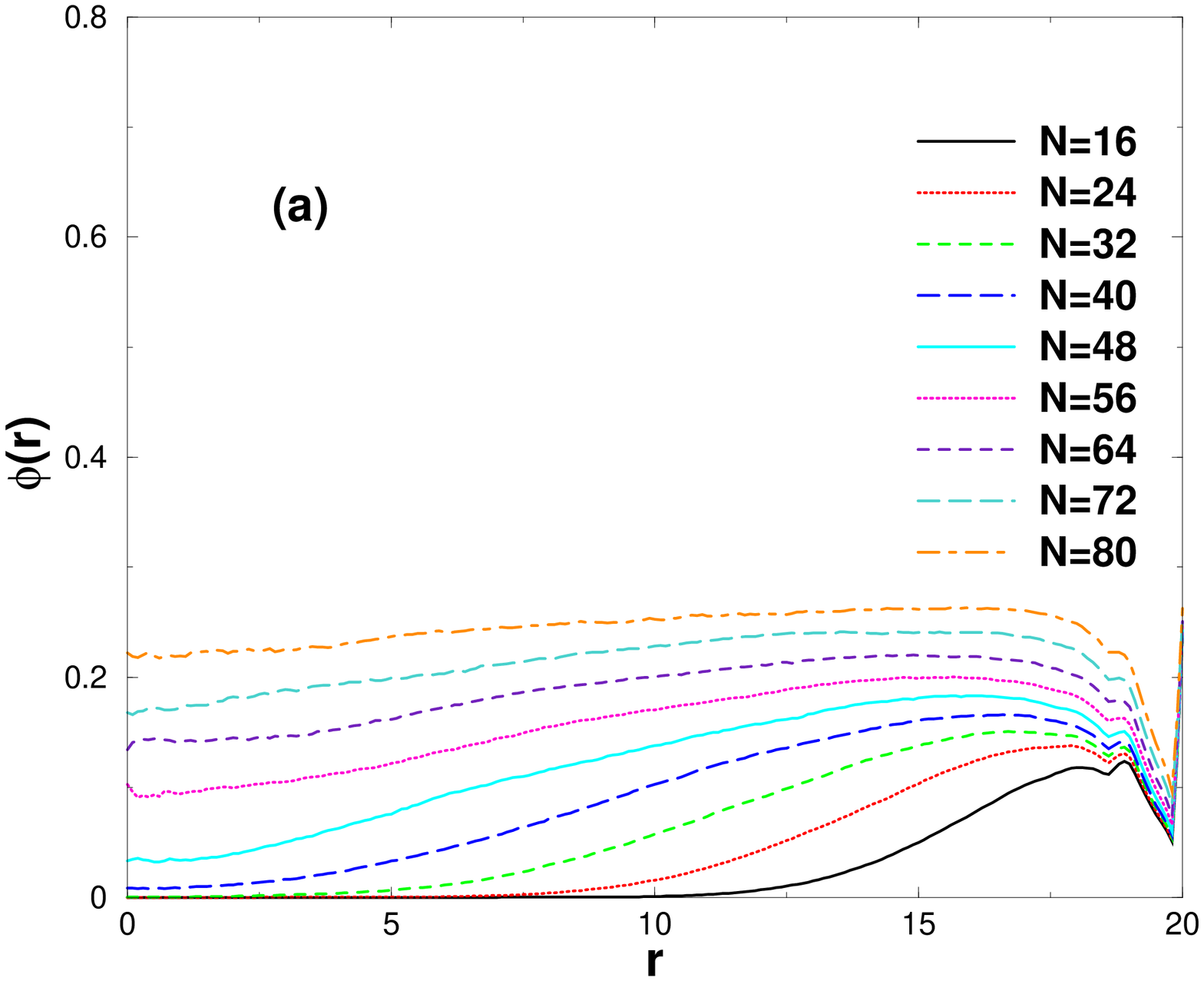}
\hspace*{50pt} \includegraphics[width=2.8in, angle=270]{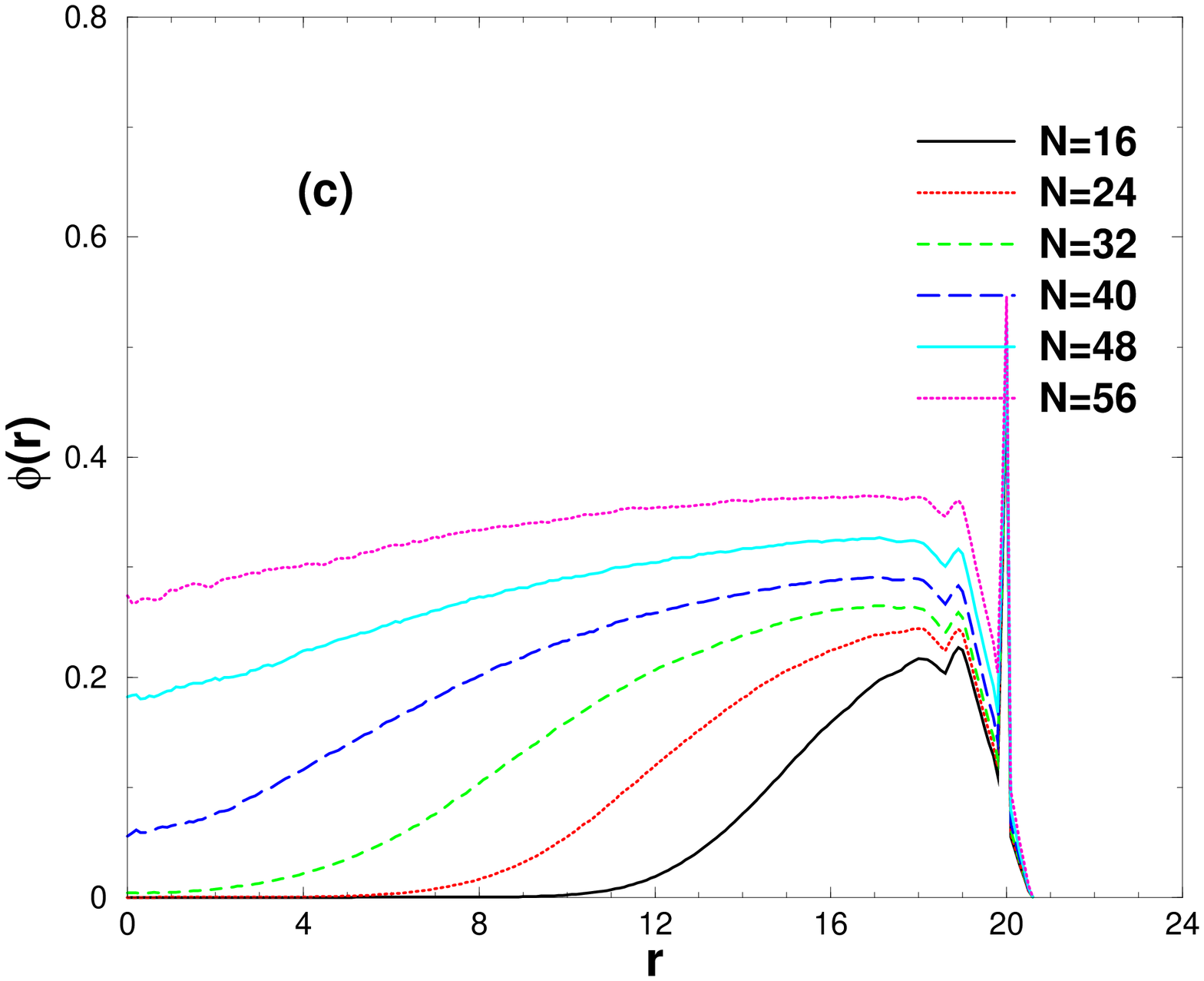}
}
\centerline{\includegraphics[width=2.8in, angle=270]{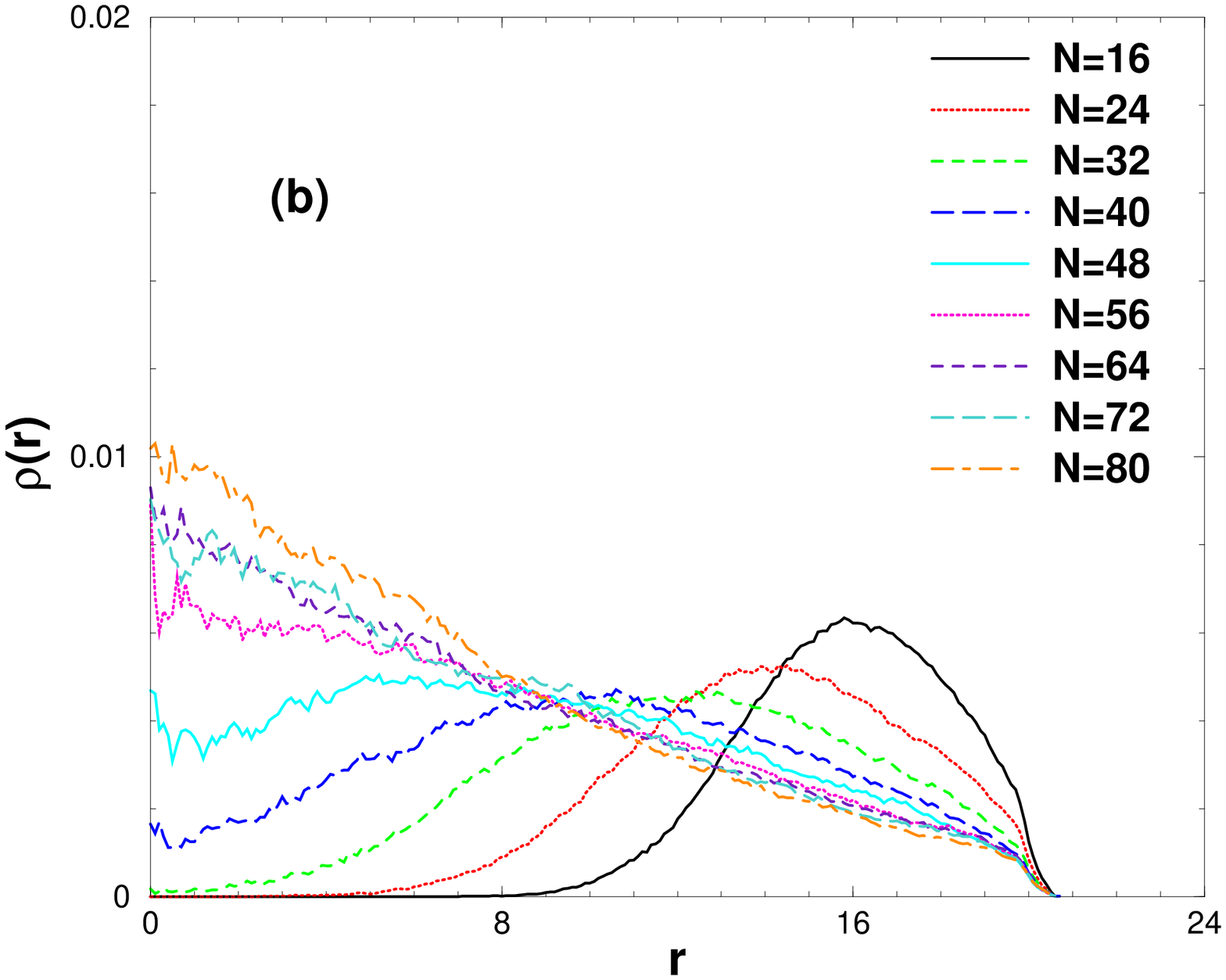}
\hspace*{50pt} \includegraphics[width=2.8in, angle=270]{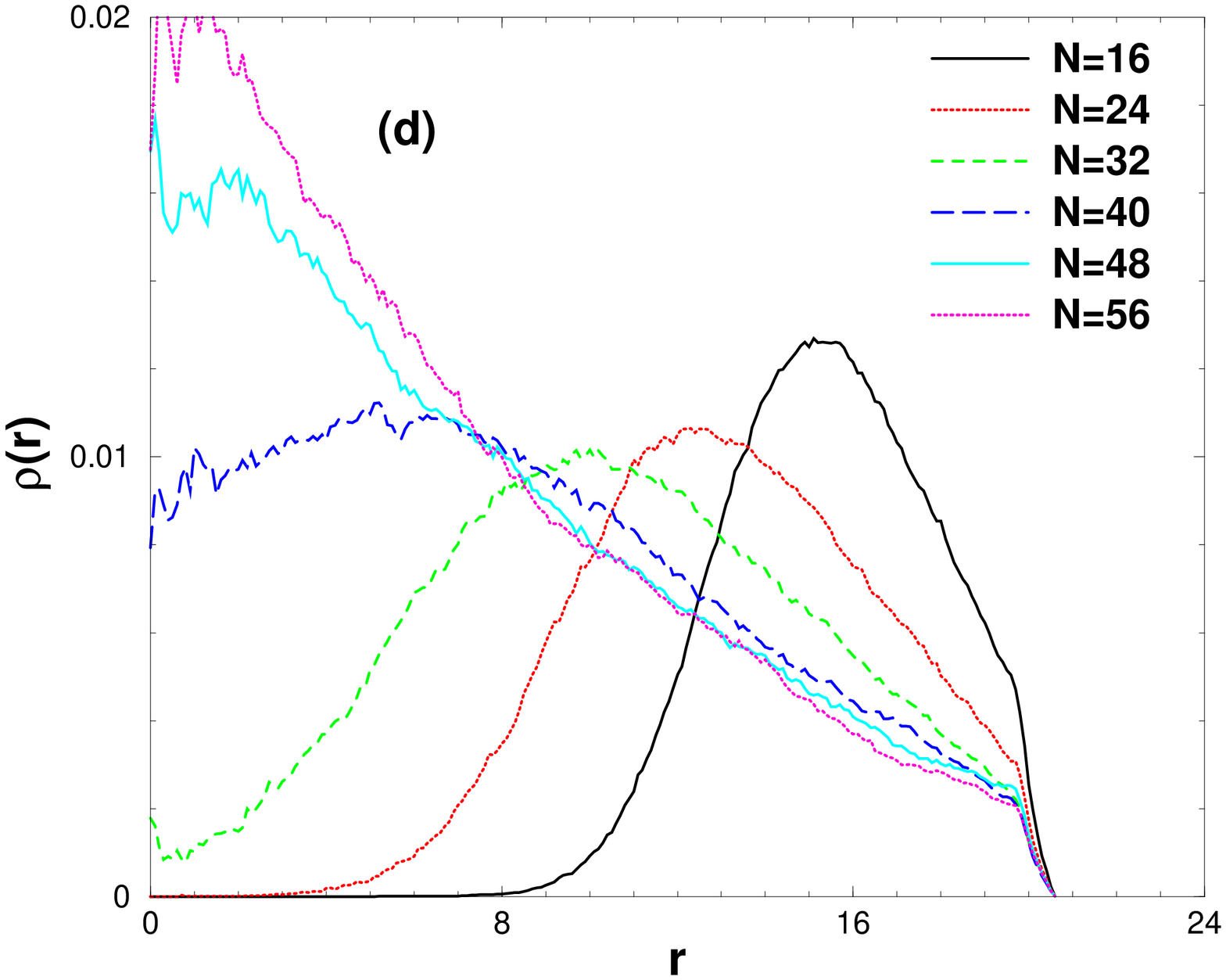}
}
\caption{Same as Fig.~\ref{fig2}, but for D=42. Part
(a,b) refers to $N_{ch}=294\; (\sigma = 0.03071)$ and part (c,d)
refers to $N_{ch} = 600 \; (\sigma = 0.06268)$\label{fig3}}
\end{figure}

\clearpage
\centerline{\includegraphics[width=2.8in, angle=270]{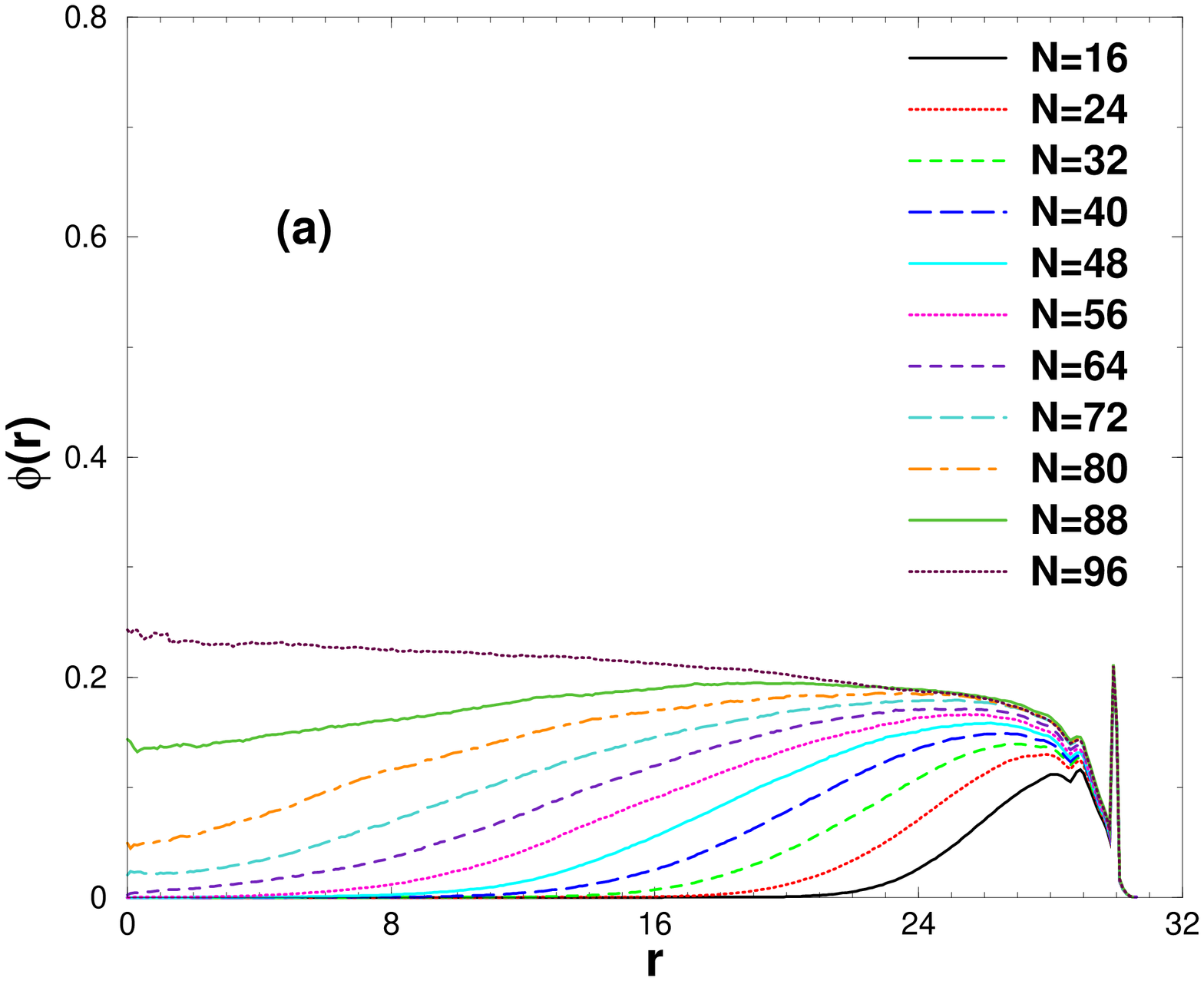}
\hspace*{50pt} \includegraphics[width=2.8in, angle=270]{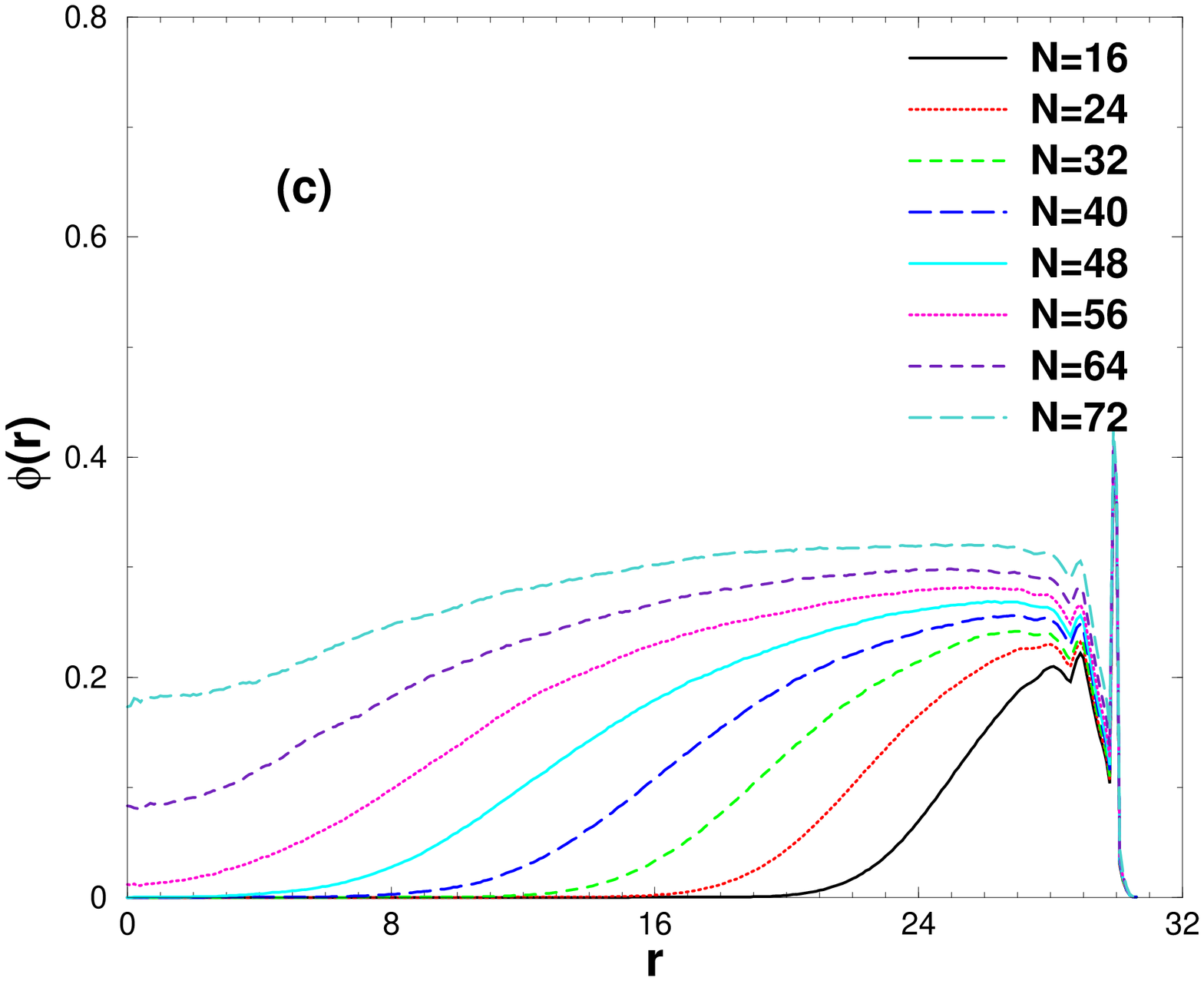}
}
\centerline{\includegraphics[width=2.8in, angle=270]{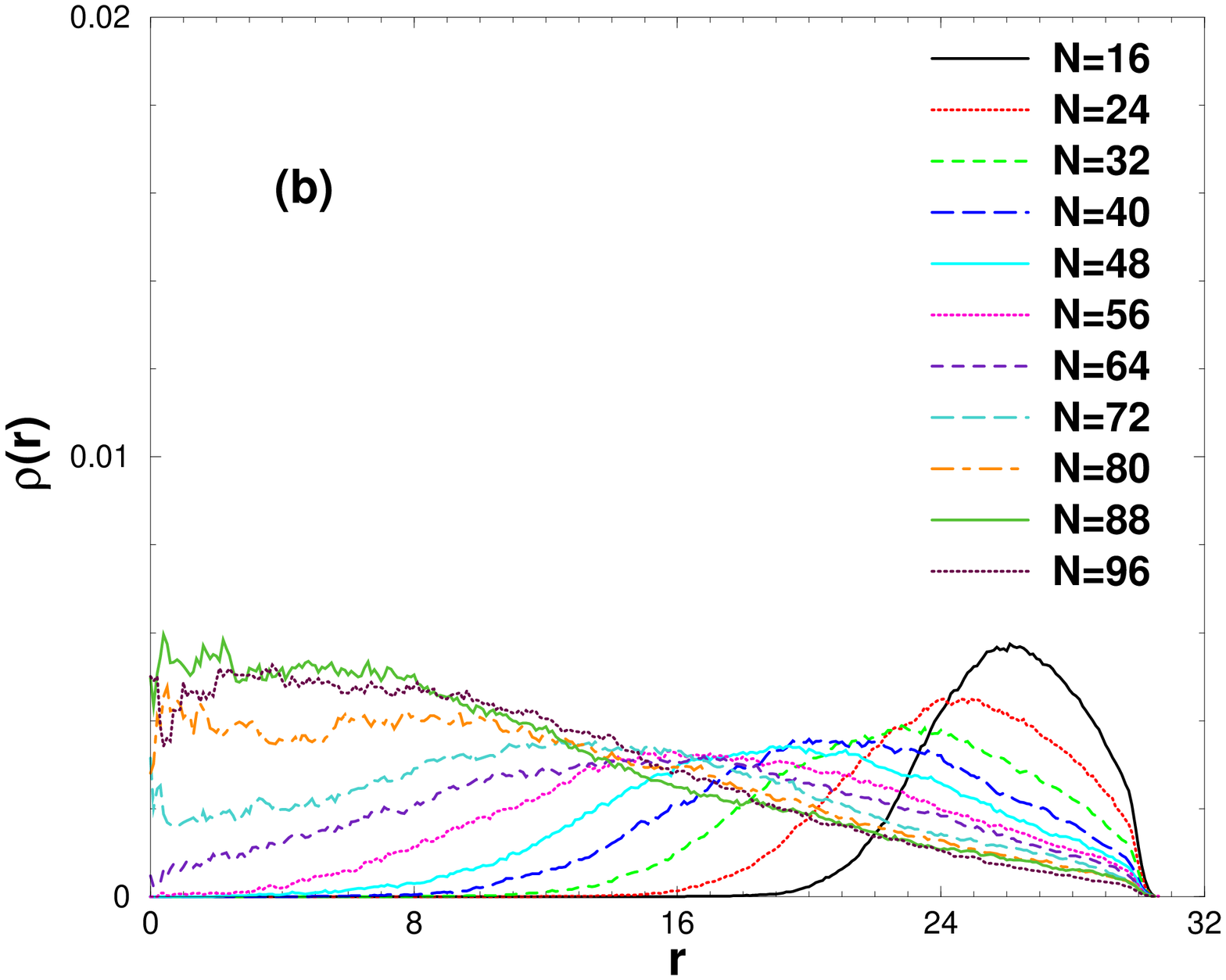}
\hspace*{50pt} \includegraphics[width=2.8in, angle=270]{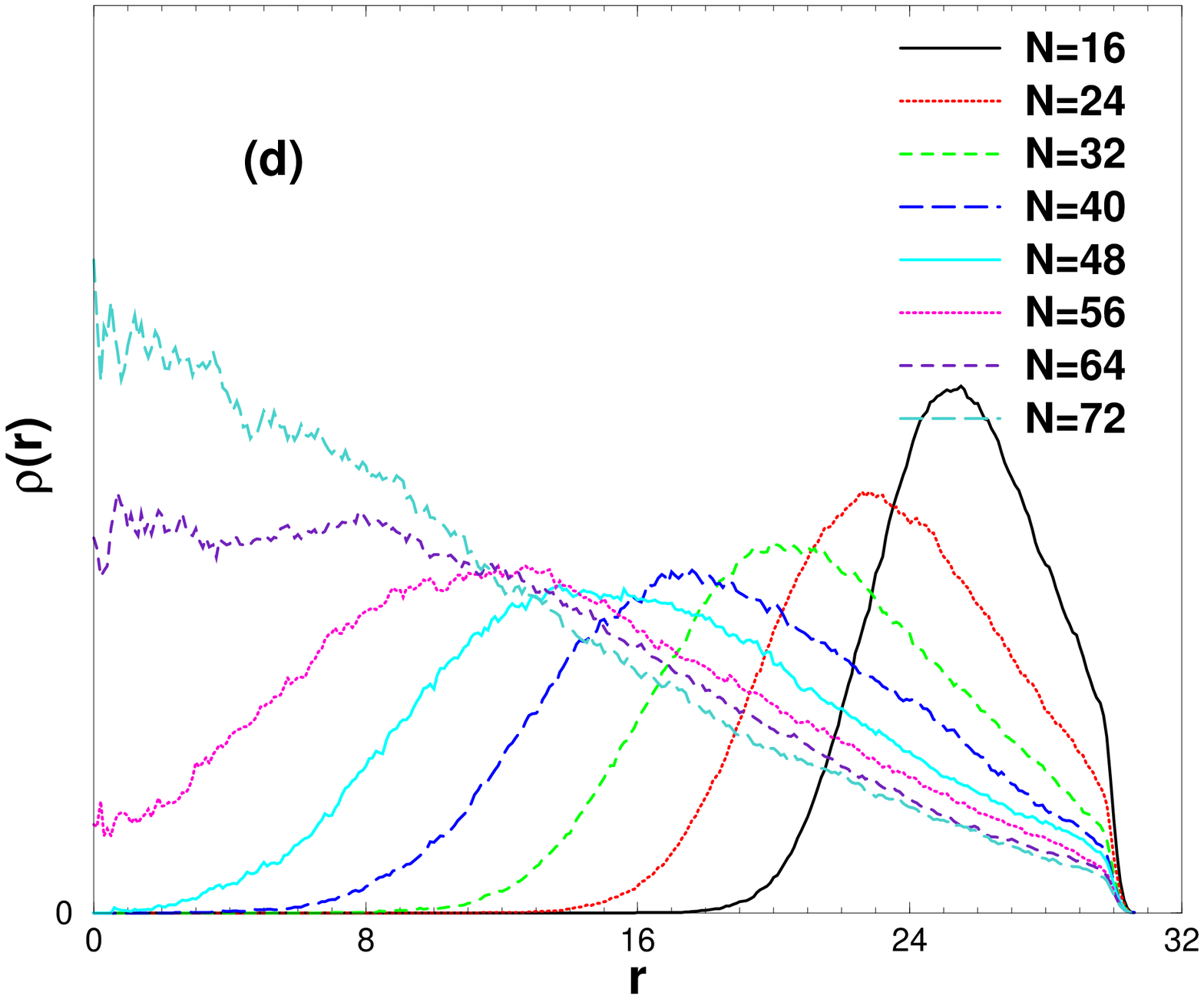}
}
\begin{figure}\caption{Same as Fig.~\ref{fig2}, \ref{fig3}, but for D=62. 
Part
(a,b) refers to $N_{ch}=384\; (\sigma = 0.03019)$ and part (c,d)
refers to $N_{ch} = 792 \; (\sigma = 0.06227)$\label{fig4}}
\end{figure}

\clearpage
\begin{figure}
\centerline{\includegraphics[width=2.8in, angle=270]{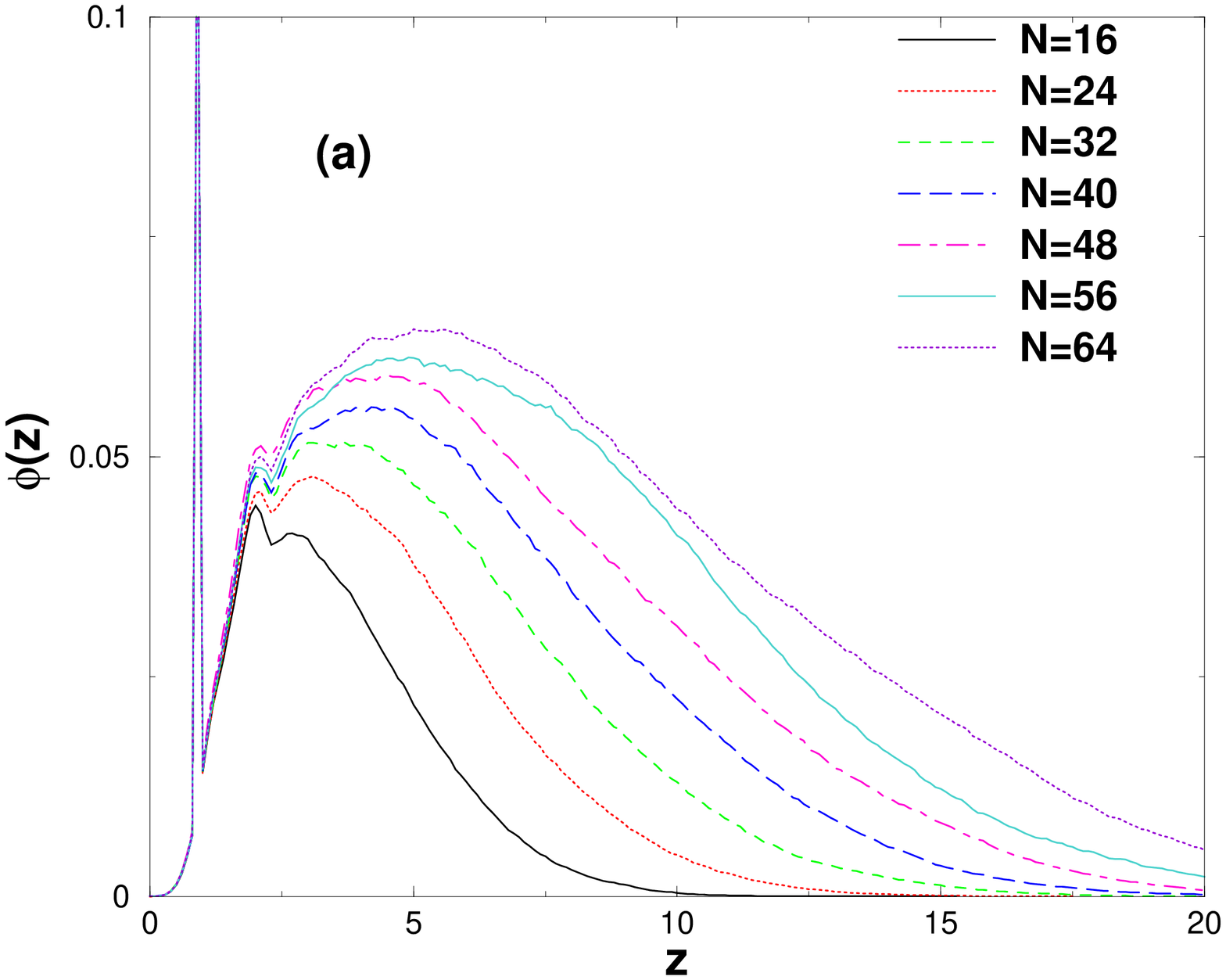}
\hspace*{20pt} \includegraphics[width=2.8in, angle=270]{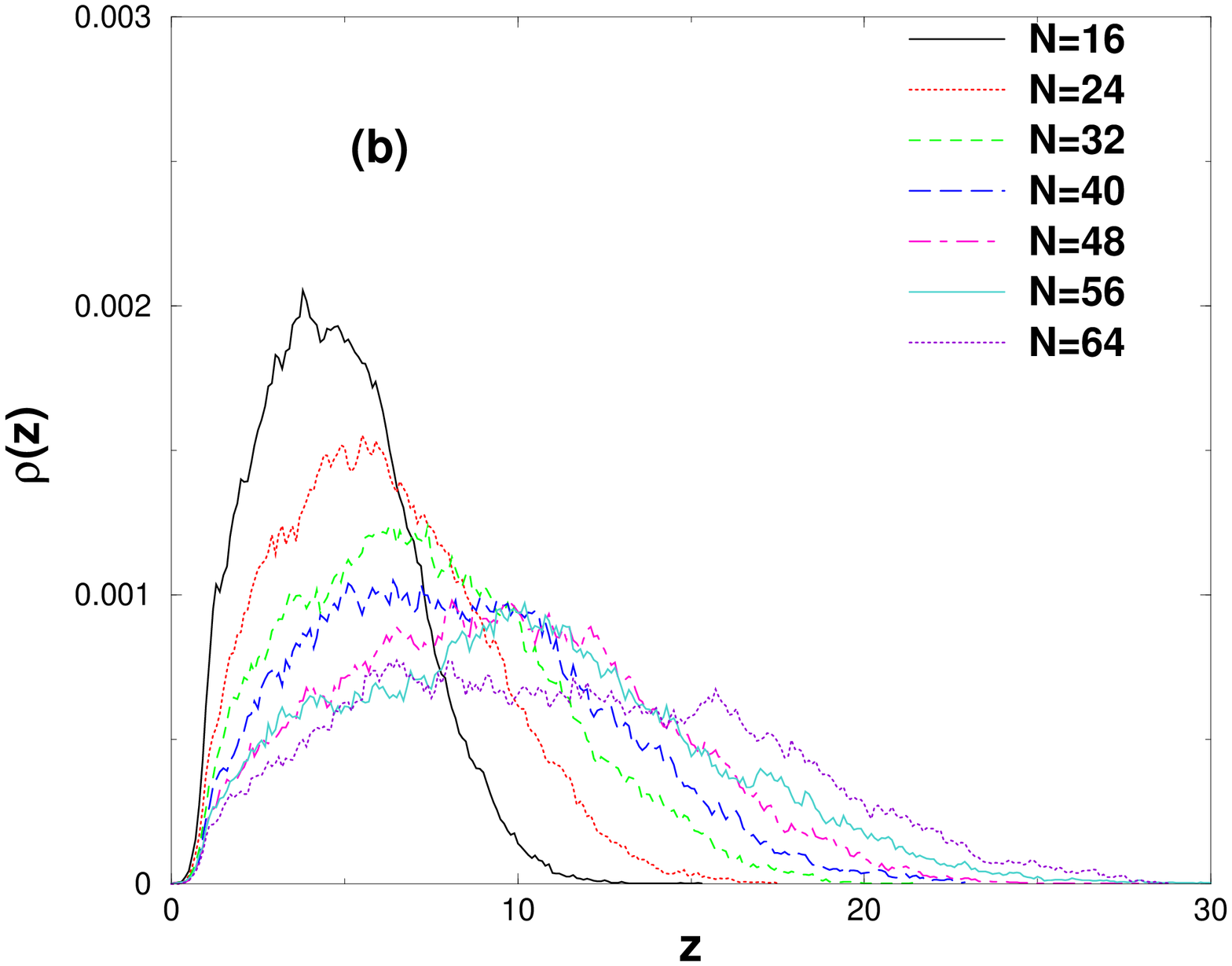}
}
\centerline{\includegraphics[width=2.8in, angle=270]{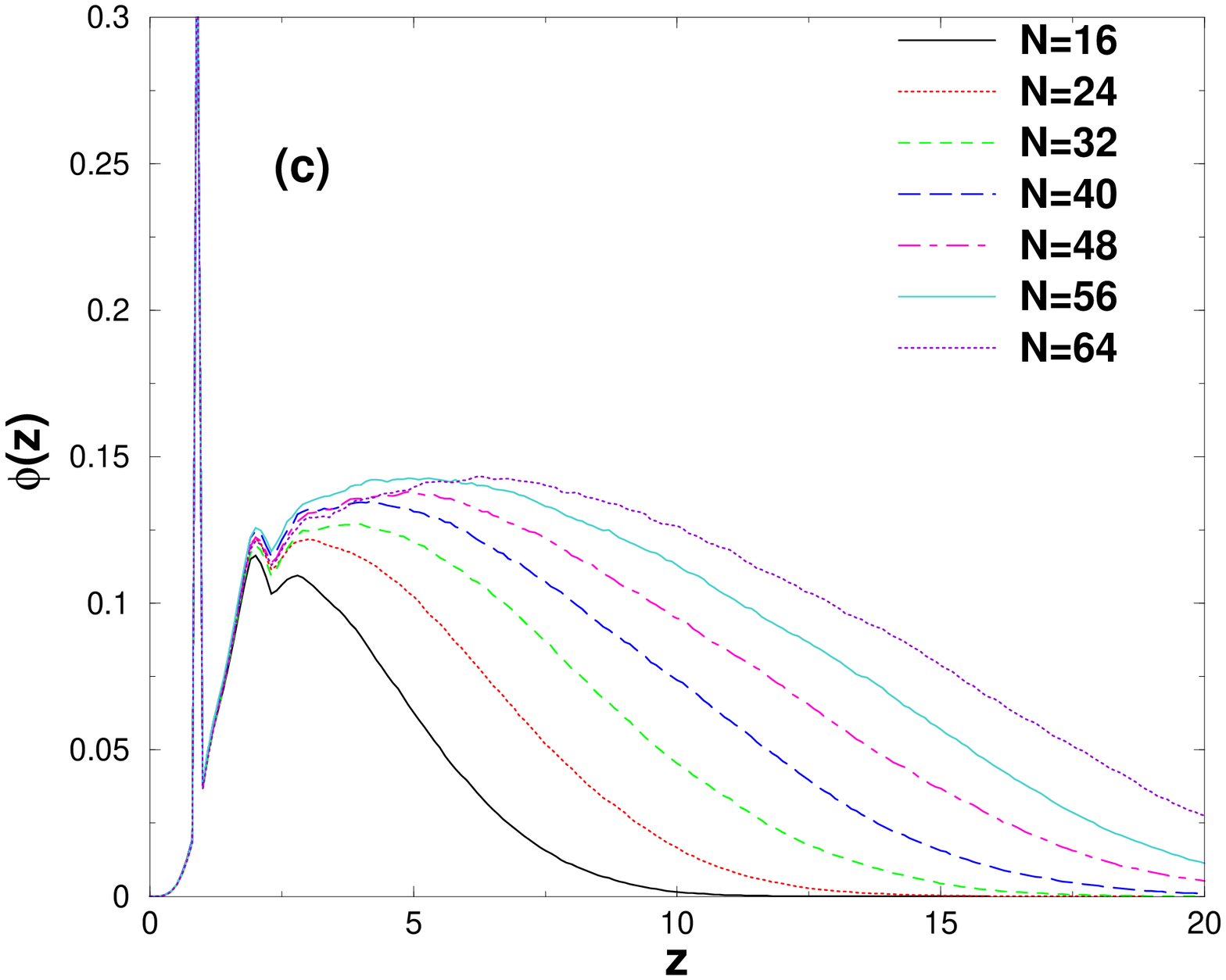}
\hspace*{20pt} \includegraphics[width=2.8in, angle=270]{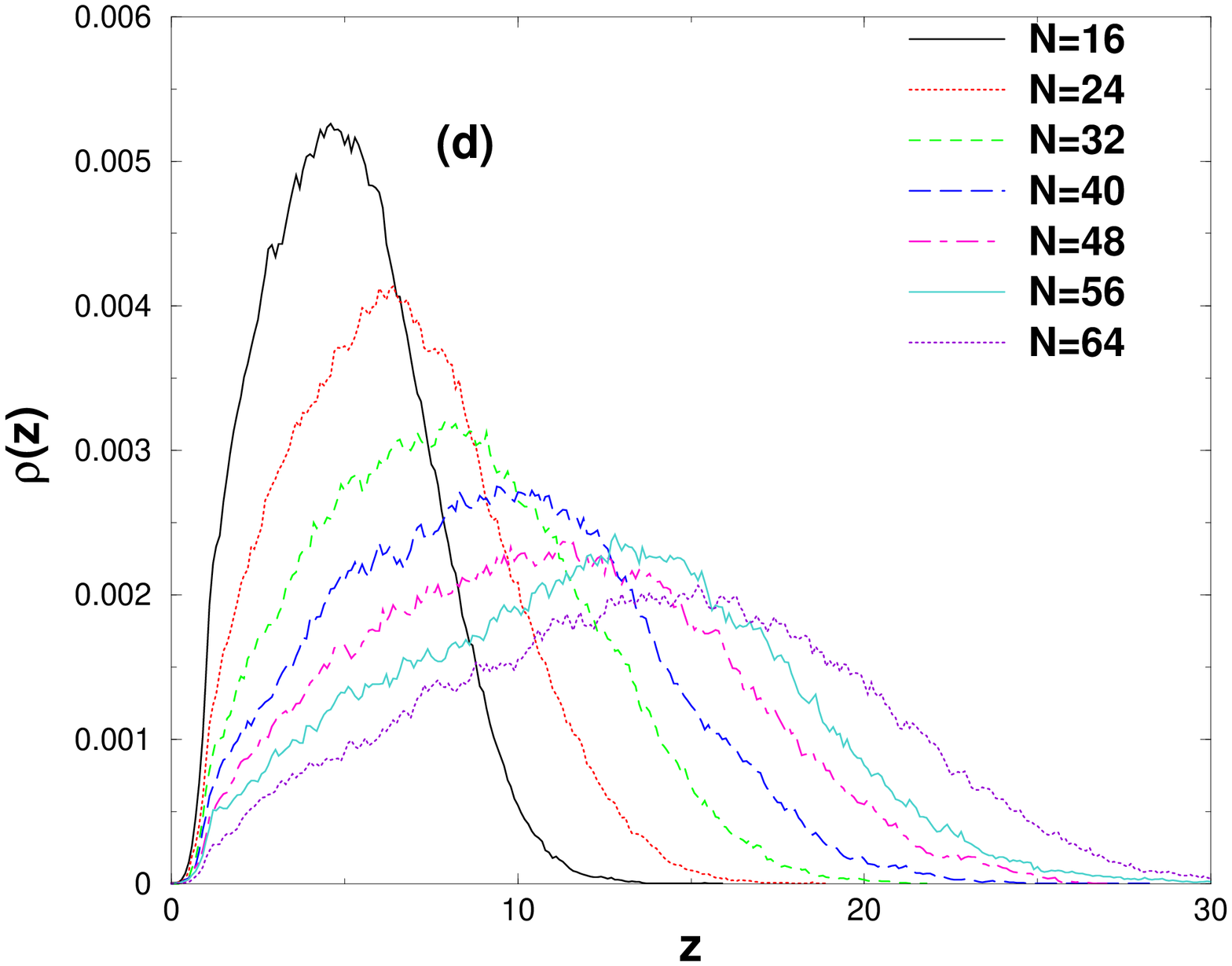}
}
\centerline{\includegraphics[width=2.8in, angle=270]{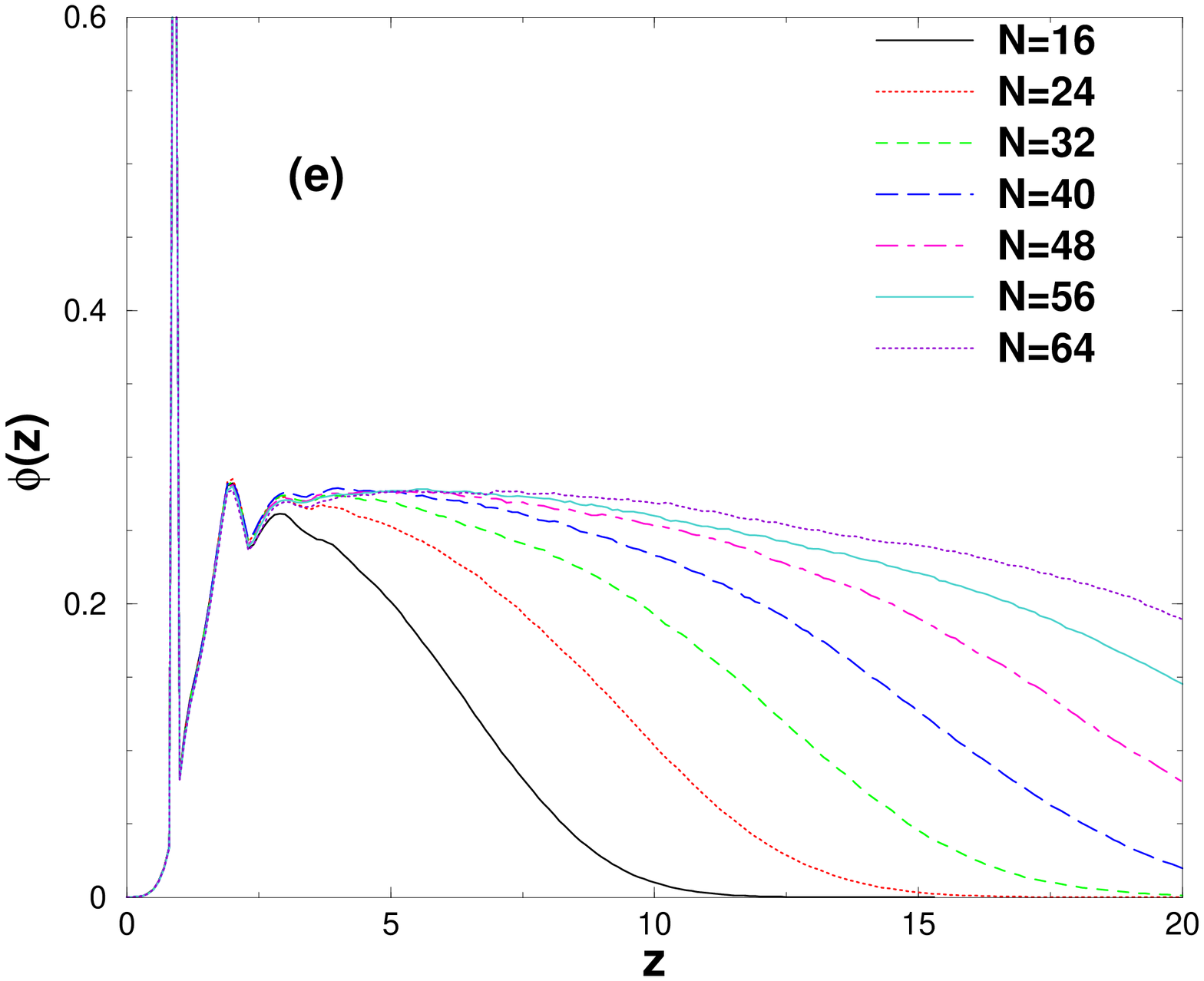}
\hspace*{20pt} \includegraphics[width=2.8in, angle=270]{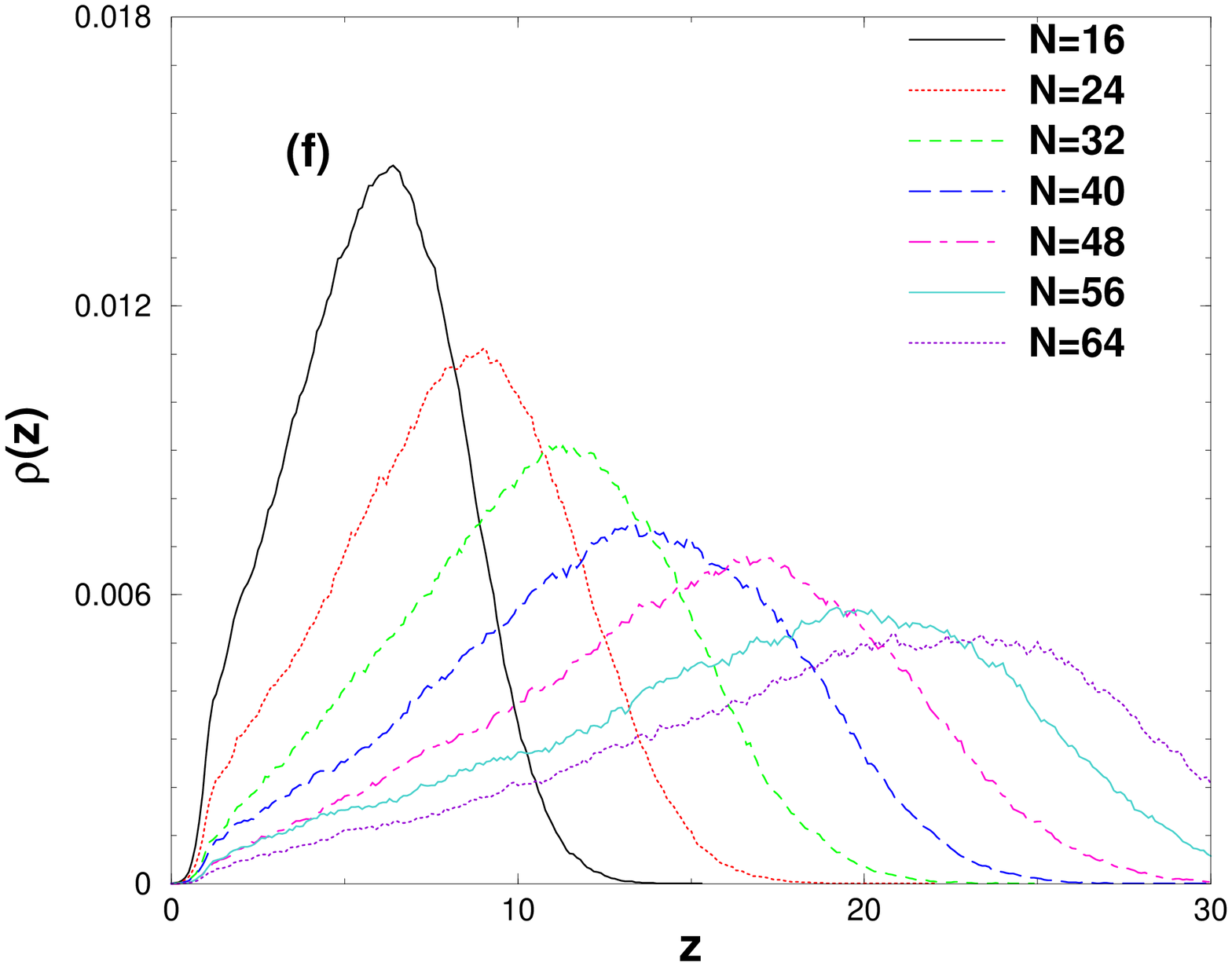}
}

\caption{Same as Fig.~\ref{fig2} - \ref{fig4}, but
for a flat substrate $(D= \infty$). Now the grafting wall is taken
at the origin of the z-axis, which runs perpendicular to the
grafting surface. Part (a,b) refers to the ``mushroom'' regime,
($N_{ch}=36 \; \sigma = 0.01155  $), while parts (c-f) refer to grafting
densities comparable to Figs.~\ref{fig2} - \ref{fig4}, namely
$N_{ch} = 100 \; \sigma = 0.03208
$, part (c,d), and $N_{ch}=289 \;
\sigma = 0.10393$, part (e,f). parts a), c), e) show the monomer densities
and parts b), d), f) the chain end distributions.
\label{fig5}}
\end{figure}

\clearpage
\centerline{\includegraphics[width=2.8in, angle=270]{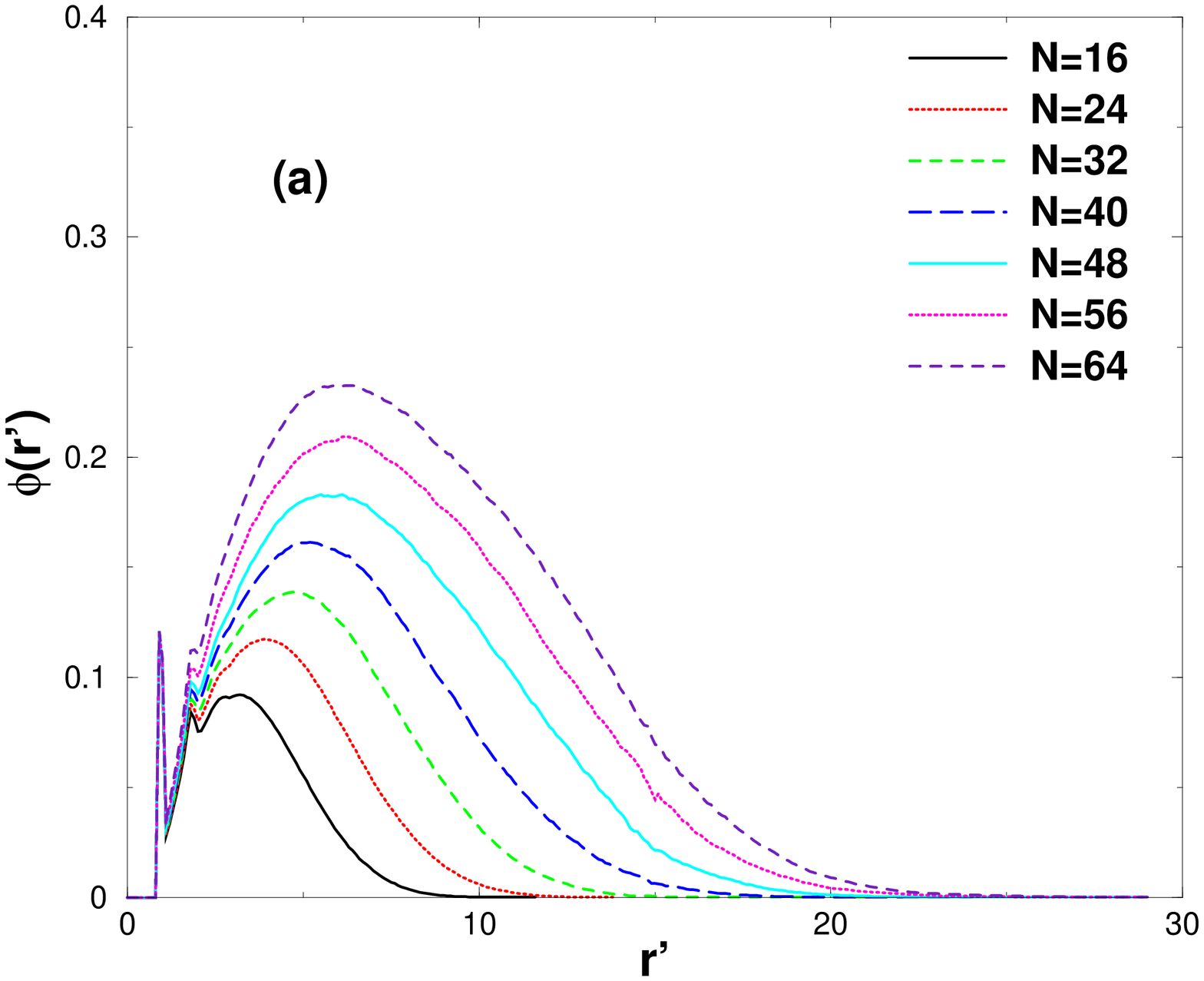}
\hspace*{50pt} \includegraphics[width=2.8in, angle=270]{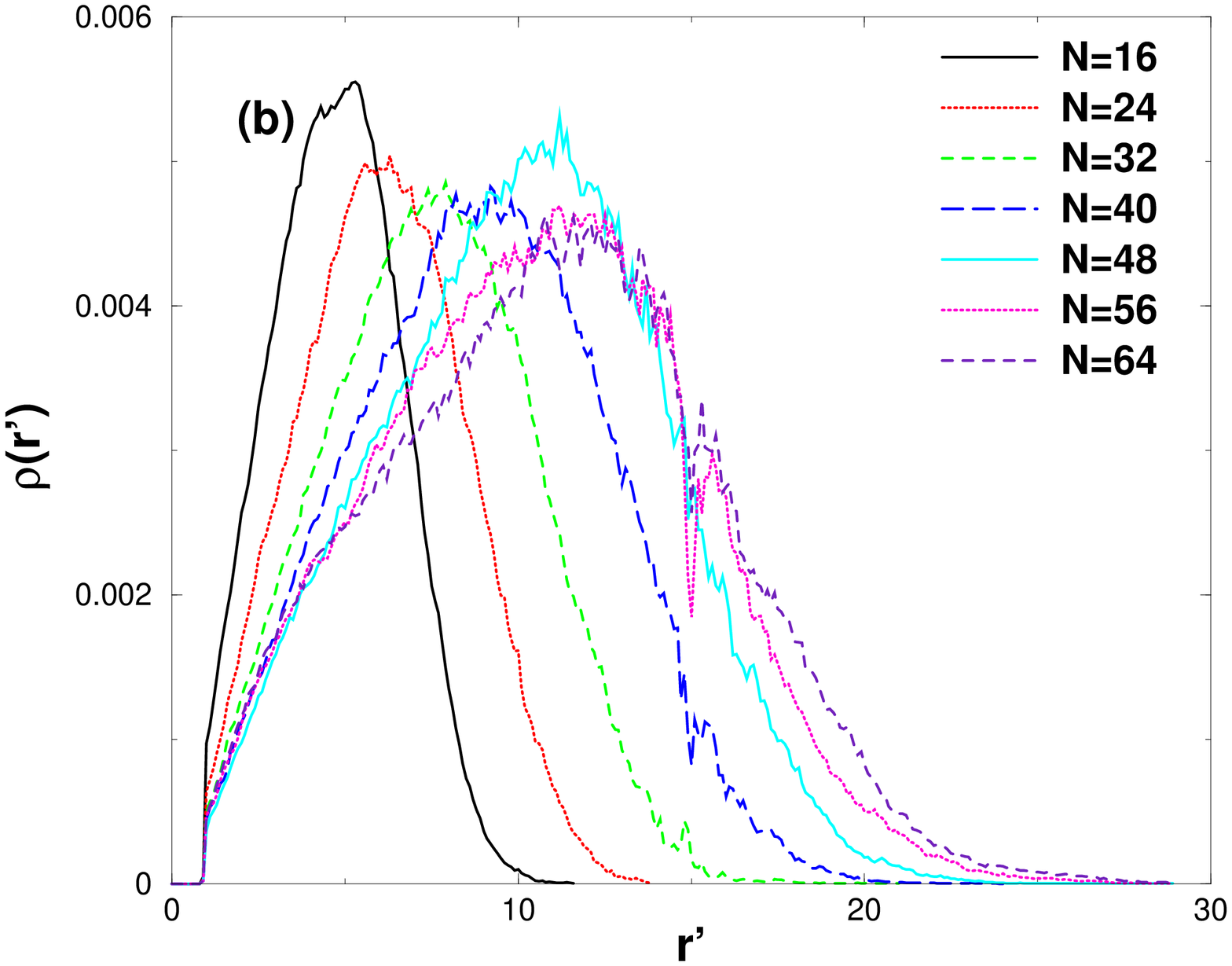}
}
\begin{figure}\caption{Distributions of all monomers of a chain,
$\phi (r')$, part a, and of chain ends, $\rho (r')$, part b, plotted
vs the distance $r'$ from the grafting wall, for a cylindrical tube
of diameter $D=30$, grafting density $\sigma = 0.01$, and various
chain lengths, as indicated in the figure. \label{fig6}}
\end{figure}

\clearpage
\begin{figure}
\centerline{\includegraphics[width=2.8in, angle=270]{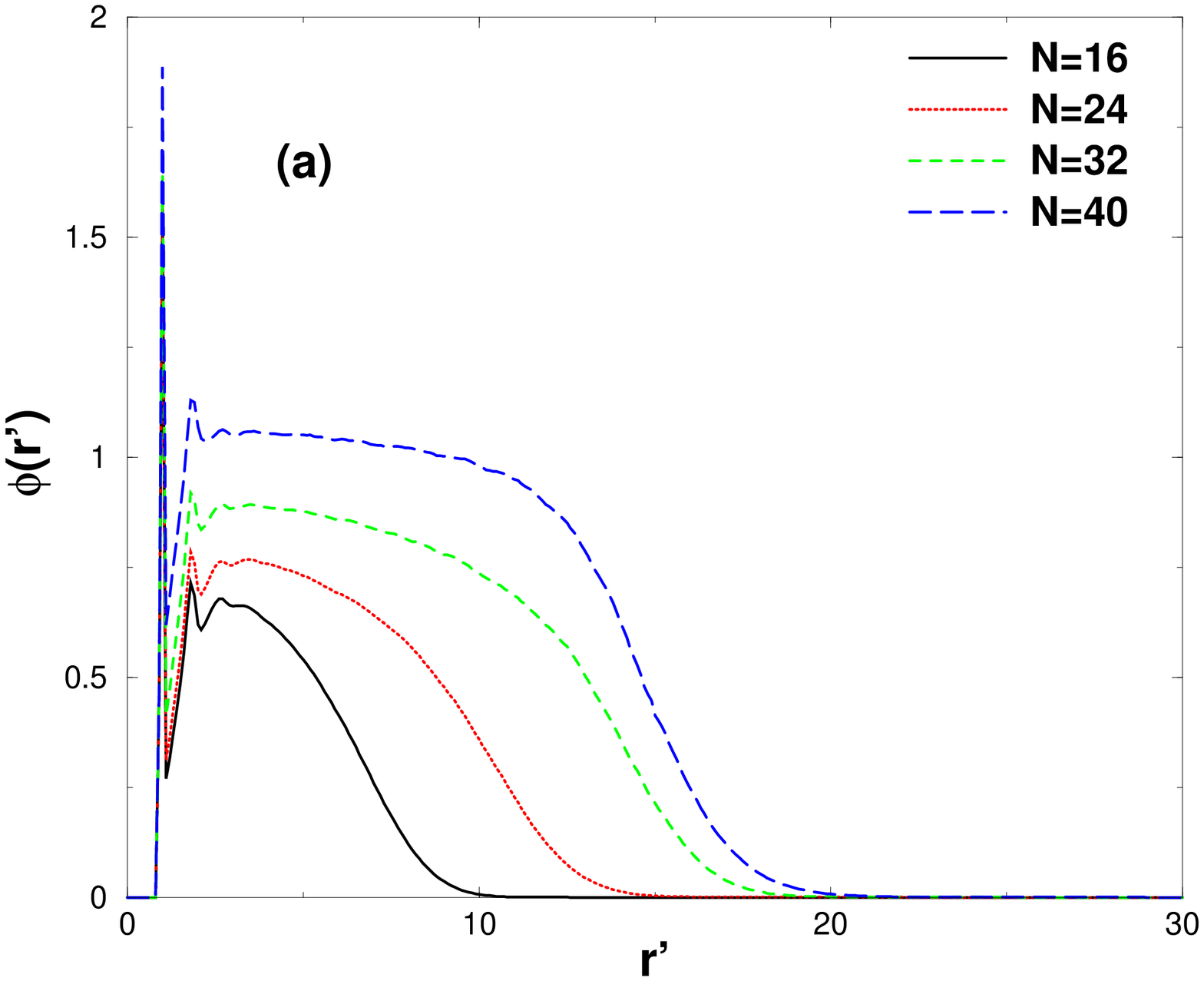}
\hspace*{50pt} \includegraphics[width=2.8in, angle=270]{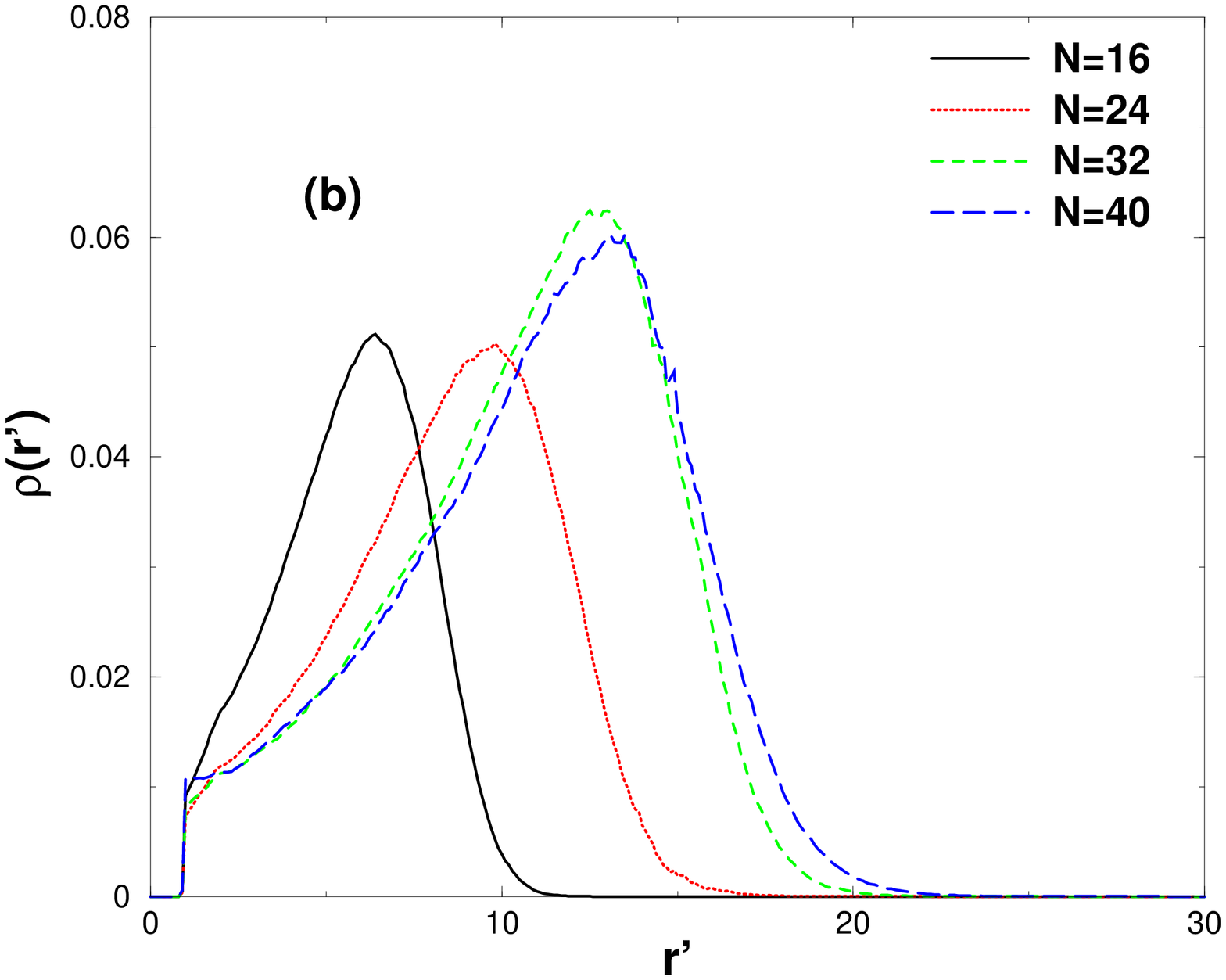}
}
\caption{Same as Fig.~\ref{fig6}, but for $\sigma =
0.09$.\label{fig7}}
\end{figure}

\clearpage
\centerline{\includegraphics[width=2.8in, angle=270]{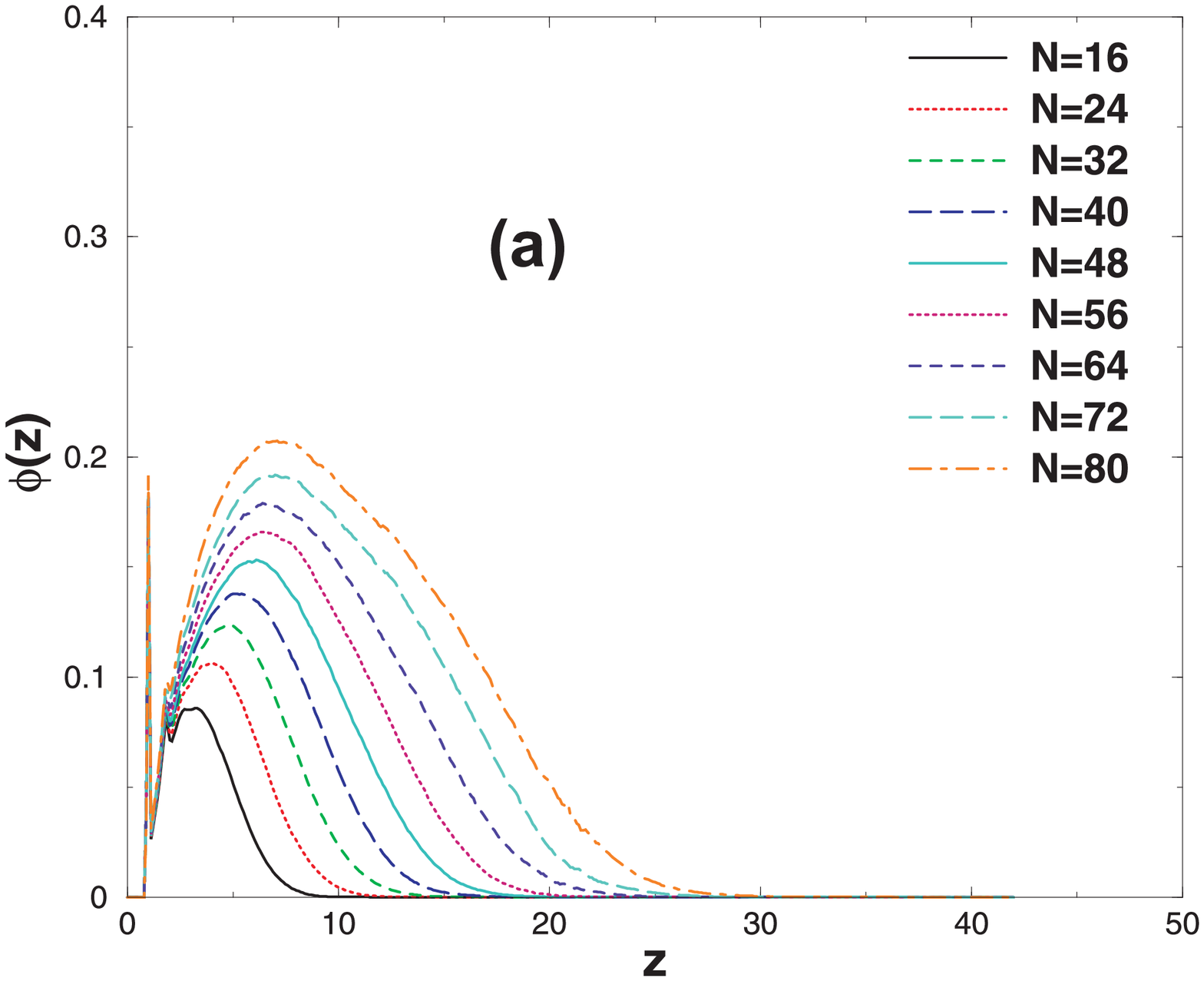}
\hspace*{50pt} \includegraphics[width=2.8in, angle=270]{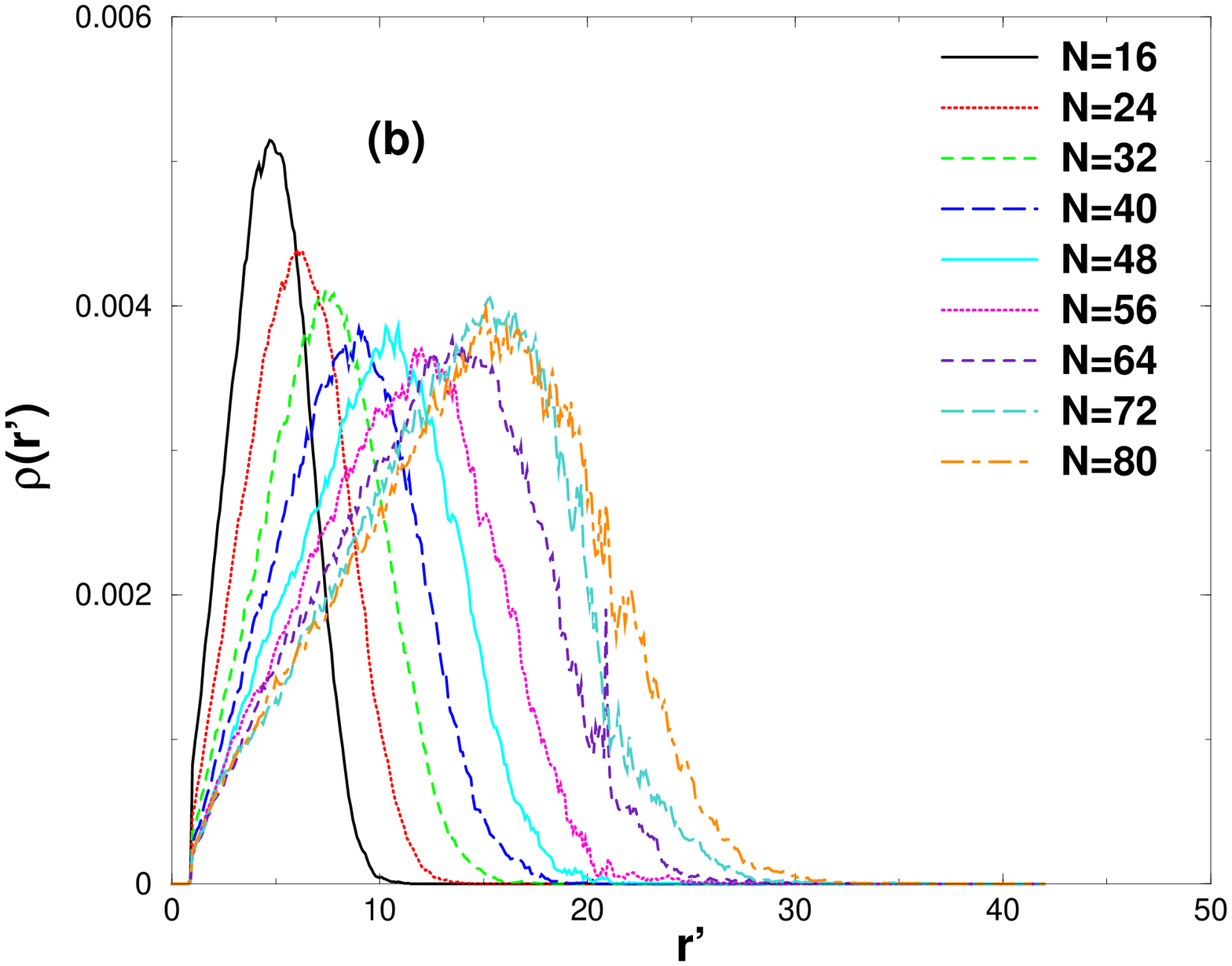}
}
\begin{figure}\caption{Same as Fig.~\ref{fig6}, but for $D=42$.\label{fig8}}
\end{figure}

\clearpage
\begin{figure}
\centerline{\includegraphics[width=2.8in, angle=270]{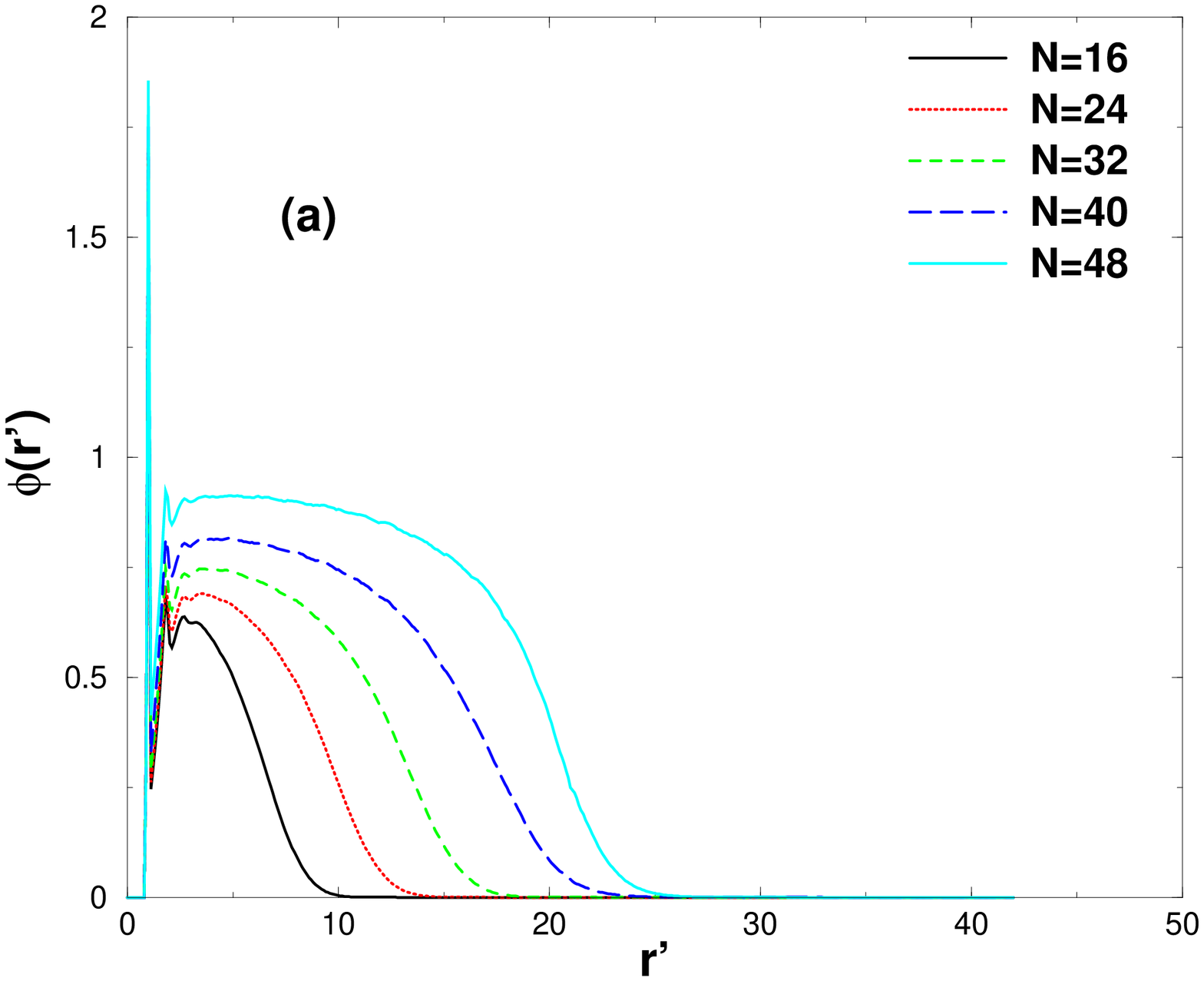}
\hspace*{50pt} \includegraphics[width=2.8in, angle=270]{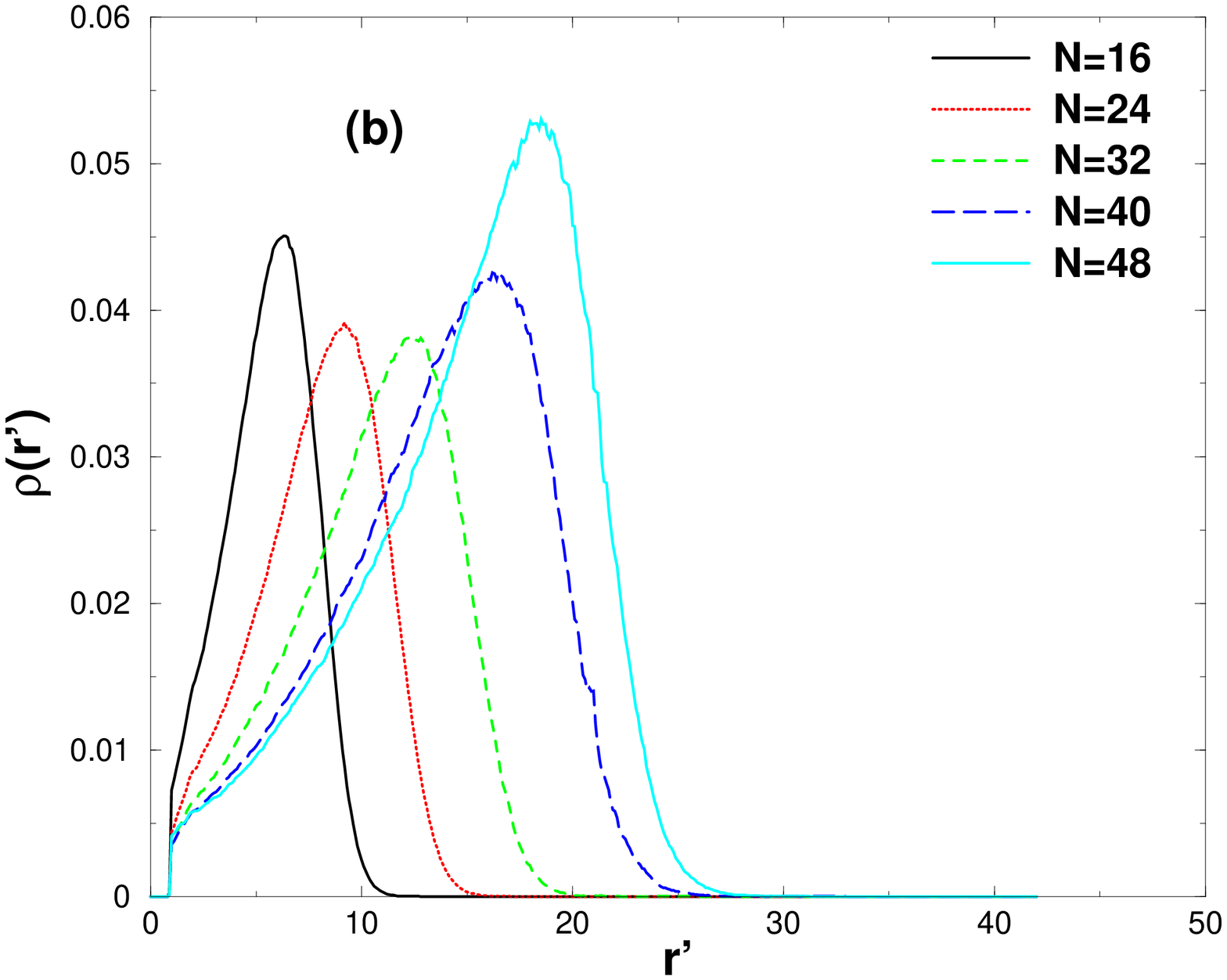}
}
\caption{Same as Fig.~\ref{fig8}, but for $\sigma =
0.09$.\label{fig9}}
\end{figure}

\clearpage
\begin{figure}
\centerline{\includegraphics[width=2.8in, angle=270]{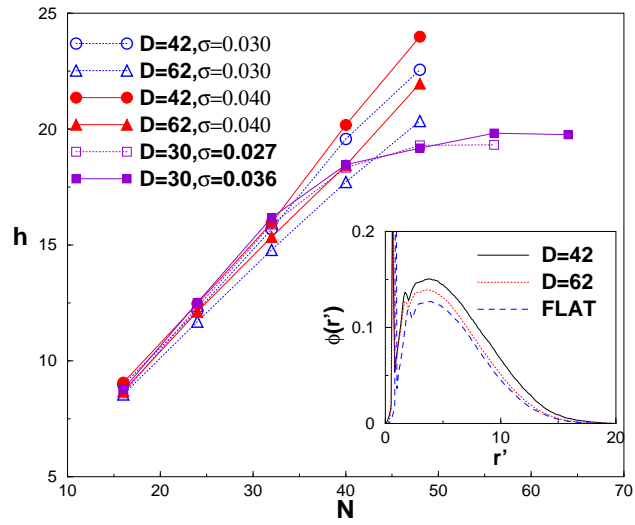}
}
\caption{Brush height $h$ plotted vs $N$ for various diameters $D$ and
grafting densities $\sigma$, as indicated.
Insert compares the density profiles $\phi(r')$ vs. $r'$ for two choices
of $D$ at $\sigma=0.03$ with the monomer density profile of a flat brush.
\label{fig10}}
\end{figure}

\clearpage
\centerline{
\includegraphics[width=5.6in, angle=0]{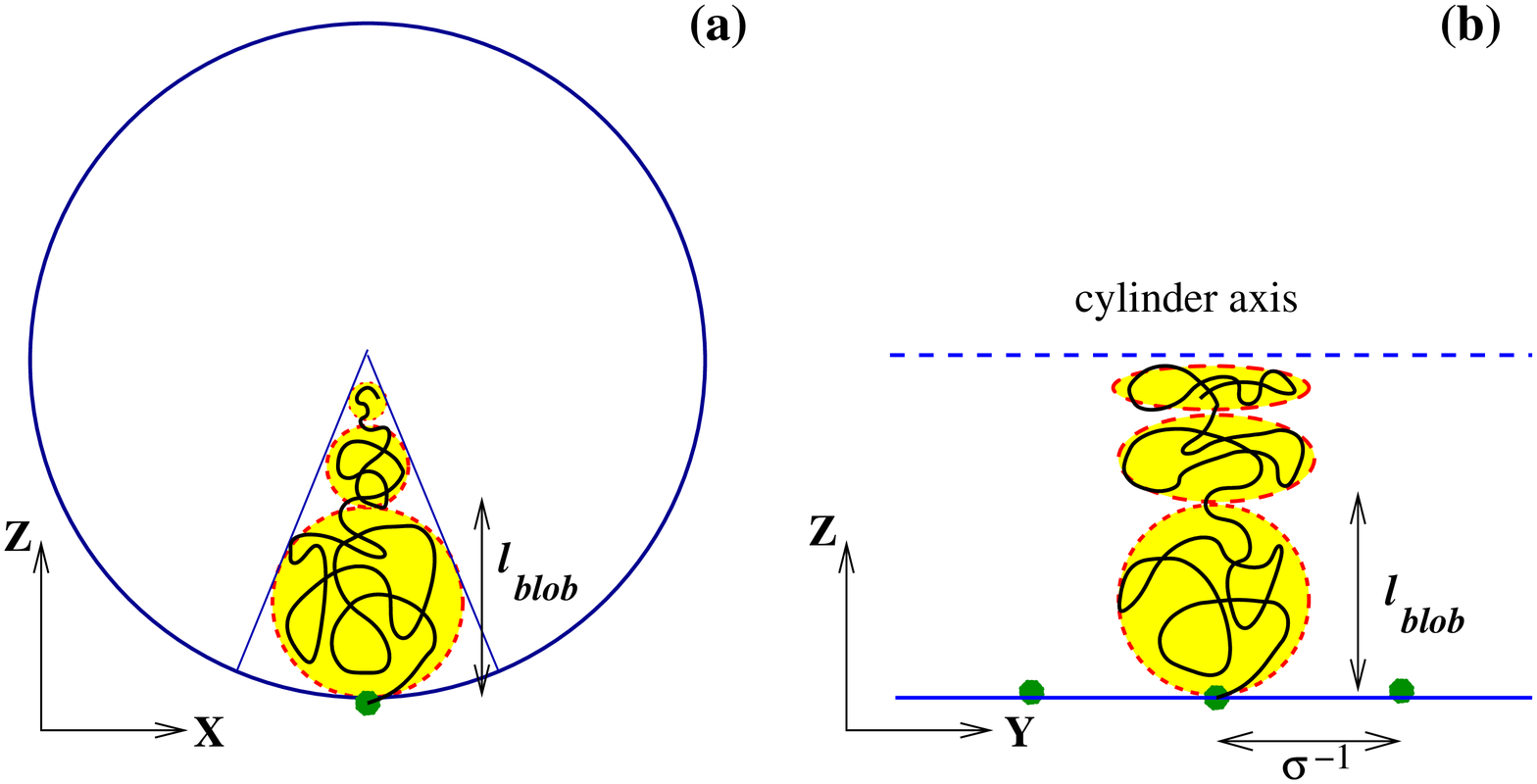}
}
\begin{figure}\caption{
Schematic illustration of the generalization of the Alexander de Gennes
blob picture for brushes to cylindrical geometry due to Sevick \cite{22}.
It is implied that each chain occupies in the $zx$-plane (cross-section of the cylinder) 
a conical sector (a), with the size $l_{blob}$ of the outermost blob being delimited
by the grafting density. In the $zy$-plane containing the cylinder axis,
all blobs have a linear dimension $l_{blob}$ along the $y$ direction, but become
progressively smaller in the $z$ direction towards the cylinder axis (b). Note that
our simulations rule out the validity of this simple picture, entropy
favors that the chains share a much larger volume.
\label{fig11}}
\end{figure}


\clearpage

\begin{thebibliography}{99}
\bibitem{1} P. G. de Gennes, \textit{J. Phys. (France)} \textbf{1976},
\textit{37}, 1445.
\bibitem{2} P. G. de Gennes, \textit{Macromolecules} \textbf{1980}, \textit{13},
1069.
\bibitem{3} S. Alexander, \textit{J. Phys. (France)} \textbf{1977}, \textit{38},
983.
\bibitem{4} A. Halperin, M. Tirrell, and T. P. Lodge, \textit{Adv. Polym.
Sci.} \textbf{1991}, \textit{100}, 31.
\bibitem{5} S. T. Milner, \textit{Science} \textbf{1991}, \textit{251}, 905.
\bibitem{6} R. C. Advincula, W. J. Brittain, K. C. Caster, and J.
R\"uhe (eds.) \textit{``Polymer Brushes''}, Wiley-VCH, Weinheim
2004.
\bibitem{7} A. N. Semenov, \textit{Sov. Phys. JETP Lett.} \textbf{1985},
\textit{61}, 733.
\bibitem{8} S. T. Milner, T. A. Witten, and M. E. Cates, \textit{Europhys.
Lett.} \textbf{1988}, \textit{5}, 413.
\bibitem{9}S. T. Milner, T. A. Witten, and M. E. Cates,
\textit{Macromolecules} \textbf{1988}, \textit{21}, 2610.
\bibitem{10} S. T. Milner and T. A. Witten, \textit{J. Phys. (France)}
\textbf{1988}, \textit{49}, 1951.
\bibitem{11} T. Cosgrove, T. Heath, B. van Lent, F. Leermakers,
and J. M. H. M. Scheutjens, \textit{Macromolecules} \textbf{1988},
\textit{20}, 1962.
\bibitem{12} A. M. Skvortsov, A. A. Gorbunov, V A. Pavlushkov, E.
B. Zhulina, O. V. Borisov, and V. A. Pryamitsin, \textit{Polym.
Sci. USSR Ser. A} \textbf{1988}, \textit{20}, 1706.
\bibitem{13} E. B. Zhulina, V. A. Pryamitsin, and O. V. Borisov,
\textit{Polymer Sci. USSR Ser. A}, \textbf{1989}, \textit{31},
205.
\bibitem{14} S. T. Milner, Z. G. Wang, and T. A. Witten,
\textit{Macromolecues} \textbf{1989}, \textit{22}, 489.
\bibitem{15} E. B. Zhulina, O. V. Borisov, and V. A. Pryamitsin,
\textit{J. Colloid Interface Sci.} \textbf{1990}, \textit{137},
495.
\bibitem{16} B. van Lent, R. Israels, J. M. H. H. Scheutjens, and
G. J. Fleer, \textit{J. Colloid Interface Sci.} \textbf{1990},
\textit{137}, 380.
\bibitem{17} R. C. Ball, J. F. Marko, S. T. Milner, and T. A.
Witten, \textit{Macromolecules} \textbf{1991}, \textit{24}, 693.
\bibitem{18} E. B. Zhulina, O. V. Borisov, V. A. Pryamitsin, and
T. M. Birshtein, \textit{Macromolecules} \textbf{1991},
\textit{24}, 140.
\bibitem{19} C. M. Wijmans, J. M. H. M. Scheutjens, and E. B.
Zhulina, \textit{Macromolecules} \textbf{1992}, \textit{25}, 2657.
\bibitem{20} C. M. Wijmans and E. B. Zhulina, \textit{Macromolecues}
\textbf{1993},
\textit{26}, 7214.
\bibitem{21} N. Dan and M. Tirrell, \textit{Macromolecules} \textbf{1993},
\textit{26}, 637.
\bibitem{22} E. M. Sevick, \textit{Macromolecules} \textbf{1996}, \textit{29},
6952.
\bibitem{23} C. Hiergeist and R. Lipowsky, \textit{J. Phys. II (France)}
\textbf{1996}, \textit{6}, 1465.
\bibitem{24} E. N. Govorun and I. Erukhimovich, \textit{Langmuir},
\textbf{1999},
\textit{15}, 8392.
\bibitem{25} M. Manghi, M. Aubouy, C. Gay and C. Ligoure, \textit{Eur.
Phys. J. E.} \textbf{2001}, \textit{5}, 519.
\bibitem{26} M. Murat and G. S. Grest, \textit{Phys. Rev. Lett.} \textbf{1989},
\textit{63},
1074.
\bibitem{27} M. Murat and G. S. Grest, \textit{Macromolecules} \textbf{1989},
\textit{22},
4054.
\bibitem{28} A. Chakrabarti and R. Toral, \textit{Macromolecules} \textbf{1990},
\textit{23},
2016.
\bibitem{29} P.-Y. Lai and K. Binder,\textit{ J. Chem. Phys.} \textbf{1991},
\textit{95},
9288.
\bibitem{30} M. Murat and G. S. Grest, \textit{Macromolecules} \textbf{1991},
\textit{24},
704.
\bibitem{31} G. S. Grest and M. Murat, \textit{Macromolecules} \textbf{1993},
\textit{26},
3108.
\bibitem{32} J. Wittmer, A. Johner, J.-F. Joanny and K. Binder, \textit{J.
Chem. Phys.} \textbf{1994}, \textit{101}, 4379.
\bibitem{33} K. Binder, P.-Y. Lai, and J. Wittmer, \textit{Faraday
Discuss.} \textbf{1994}, \textit{98}, 97.
\bibitem{34} G. S. Grest and M. Murat, in: \textit{Monte Carlo and
Molecular Dynamics Simulations in Polymer Science,} edited by K.
Binder, Oxford Univ. Press, New York \textbf{1995}, p. 476.; G. S. Grest, 
\textit{Adv. Polym. Sci}. \textbf{1999}, \textit{138},
149.

\bibitem{35} K. Prochazka, \textit{J. Phys. Chem.} \textbf{1995}, \textit{99},
14108.
\bibitem{36} I. Szleifer and M. Carignano, \textit{Adv. Chem. Phys.}. \textbf{1996}, \textit{94},
165; I. Szleifer and M. Carignano, \textit{J. Chem. Phys.}. \textbf{1995}, \textit{102},
8662.
\bibitem{37} P. Sotta, A. Lesne and J. M. Victor, \textit{J. Chem. Phys}.
\textbf{2000}, \textit{112}, 1565.
\bibitem{38} T. Kreer, S. Metzger, M. M\"uller, K. Binder and J.
Baschnagel, \textit{J. Chem. Phys.} \textbf{2004}, \textit{120},
4012.
\bibitem{39} F. Tessier and G. W. Slater, \textit{Macromolecules} \textbf{2005},
\textit{38}, 6752; \textbf{2006}, \textit{39}, 1250.
\bibitem{40} K. Binder, in: \textit{Computational Modelling of Polymers},
edited by J. Bicerano, Marcel Dekker, New York, 1992, p. 221.
\bibitem{41} K. Binder, ed. ``\textit{Monte Carlo and Molecular Dynamics
of Condensed Matter Systems}'', Societa Italiana di Fisica,
Bologna 1996.
\bibitem{42} K. Binder and G. Ciccotti, eds. ``\textit{Monte Carlo and
Molecular Dynamics of Condensed Matter Systems}'', Societa
Italiana di Fisica, Bologna, 1996.
\bibitem{43} K. Binder and A. Milchev, \textit{J. Computer-Aided Mater.
Design} \textbf{2000}, \textit{9}, 93.
\bibitem{44} M. Kotelyanskii and D. N. Theodorou, eds. ``\textit{Computer
Simulation Methods for Polymers}'', Marcel Dekker, New York, 2004.
\bibitem{45} T. Kreer, M. H. M\"user, K. Binder, and J. Klein,
\textit{Langmuir} \textbf{2001}, \textit{17}, 7804.
\bibitem{46} P. J. Flory, ``\textit{Principles of Polymer Chemistry}'',
Cornell University Press, Ithaca, N. Y., 1953.
\bibitem{47} M. M\"uller and L.-G. MacDowell, \textit{Macromolecules}
\textbf{2000},
\textit{33}, 3902.
\bibitem{48} G. S. Grest and K. Kremer, \textit{Phys. Rev. A} \textbf{1986},
\textit{33},
3628.
\bibitem{49} C. Bennemann, K. Binder, W. Paul, and B. D\"unweg,
\textit{Phys. Rev. E} \textbf{1998}, \textit{57}, 843.
\bibitem{50} L. Wenning, M. M\"uller, and K. Binder, \textit{Europhys.
Lett.} \textbf{2005}, \textit{71}, 639.
\bibitem{51} M. Meyyappan, ed. ``\textit{Carbon Nanotubes: Science and
Applications}'' CRC Press, Boca Raton, 2004
\bibitem{52} Z. N. Yu, H. Gao, W. Wu, H.X.Ge, and S. Y. Chou, \textit{J.
Vac. Sci. Technol. B} \textbf{2003}, \textit{21}, 2874.
\bibitem{53} W. Reisner, K. J. Morton, R. R\"uhn, Y. M. Wang, Z.
Yu, M. Rosen, J. C. Sturm, S. Y. Chou, E. Frey, and R. H. Austin,
\textit{Phys. Rev. Lett.} \textbf{2005}, \textit{94}, 196101.
\bibitem{54} D. I. Dimitrov, A. Milchev, and K. Binder, J.Chem.Phys.(in
press)
\end{thebibliography}
\end{document}